# Atomically-precise synthesis and simultaneous integration of 2D transition metal dichalcogenides enabled by nano-confinement

Ce Bian[1,2,#], Yifan Zhao[3,4,#], Roger Guzman[2,#], Hongtao Liu[1,#], Hao Hu[5], Qi Qi[1,2], Ke Zhu[1,2], Hao Wang[1,2], Kang Wu[1,2], Hui Guo[1], Wanzhen He[6], Zhaoqing Wang[2], Peng Peng[3,7], Zhiping Xu[6], Wu Zhou[2,*], Feng Ding[3,4,7*], Haitao Yang[1,2,*] and Hong-Jun Gao[1,2,*]

[1] *Beijing National Center for Condensed Matter Physics and Institute of Physics, Chinese Academy of Sciences, Beijing, China*

[2] *School of Physical Sciences, University of Chinese Academy of Sciences, Beijing, China*

[3] *Suzhou Laboratory, Suzhou, China*

[4] *Institute of Technology for Carbon Neutrality, Shenzhen Institute of Advanced Technology, Chinese Academy of Sciences, Shenzhen, China*

[5] *Frontier Institute of Science and Technology, Xi'an Jiaotong University, Xi'an, China*

[6] *Department of Engineering Mechanics and Center for Nano and Micro Mechanics, Tsinghua University, Beijing, China*

[7] *Faculty of Materials Science and Energy Engineering, Shenzhen University of Advanced Technology, Shenzhen, China*

#These authors contributed equally: Ce Bian, Yifan Zhao, Roger Guzman, Hongtao Liu

*E-mail: wuzhou@ucas.ac.cn; dingf@szlab.ac.cn; htyang@iphy.ac.cn; hjgao@iphy.ac.cn

**Two-dimensional (2D) materials, such as graphene, transition metal dichalcogenides (TMDs), and hBN, exhibit intriguing properties that are sensitive to their atomic-scale structures and can be further enriched through van der Waals (vdW) integration. However, the precise synthesis and clean integration of 2D materials remain challenging. Here, using graphene or hBN as a vdW capping layer, we create a nano-confined environment that directs the growth kinetics of 2D TMDs (e.g., $NbSe_2$ and $MoS_2$), enabling precise formation of TMD monolayers with tailored morphologies, from isolated monolayer domains to large-scale continuous films and intrinsically-patterned rings. Moreover, Janus S-Mo-Se monolayers are synthesized with atomic precision via vdW-protected bottom-plane chalcogen substitution. Importantly, our approach simultaneously produces ultraclean vdW interfaces. This in situ encapsulation reliably preserves air-sensitive materials, as evidenced by the enhanced superconductivity of nano-confined $NbSe_2$ monolayers. Altogether, our study establishes a versatile platform for the controlled synthesis and integration of 2D TMDs for advanced applications.**

Two-dimensional (2D) transition metal dichalcogenides (TMDs) and their homo-/heterostructures exhibit a plethora of unique physical and chemical properties that enable emerging applications in next-generation electronics[1-5], twistronics[6-9], energy harvesting[10], sensing[11], and catalysis[12]. However, the intriguing properties of 2D TMDs are dictated by their atomic-scale structures, particularly the layer number and the atomic compositions and arrangements. For example, $MoS_2$ is a direct bandgap semiconductor only in the monolayer form, whereas multilayer $MoS_2$ undergoes a direct-to-indirect bandgap transition owing to interlayer van der Waals (vdW) coupling[13]. Moreover, monolayer $NbSe_2$, a 2D superconductor, exhibits an extremely high in-plane critical field (~32 T) originating from robust Ising pairing protected by spin-orbit interaction[14]. With the number of $NbSe_2$ layers increasing, the vdW coupling disturbs the perfect Ising pairing and introduces orbital effect, both of which diminish the critical field[14]. At an even finer level of structural control, sandwiching the transition metal atoms between different chalcogen species (i.e., S-Mo-Se) breaks the out-of-plane symmetry and creates a polar Janus TMD monolayer[15-20]. This symmetry breaking leads to a range of intriguing properties (e.g., piezoelectricity[21], thermoelectricity[22], and pyroelectricity[23]), positioning Janus TMD monolayers as a versatile platform for 2D multi-functional devices.

The sensitivity of the intriguing properties of 2D TMDs to their atomic-scale structures necessitates atomically-precise synthesis. For monolayer TMDs, precise control over the layer thickness requires effective suppression of adlayer nucleation, which remains challenging owing to a delicate balance between precursor deposition and surface diffusion (Supplementary Note S1)[24,25]. The synthesis of Janus TMD monolayers, typically through one-side substitution of their classical counterparts, has not yet achieved atomic precision, as the opposite chalcogen plane can be unintentionally modified due to poorly controlled substitution kinetics (Supplementary Note S2)[16]. Furthermore, the application of 2D TMDs often relies on their clean vdW integration with other layered materials (e.g., graphene and hBN) to preserve inherent properties[14] or enable emergent phenomena[7]. However, conventional transfer methods for fabricating vdW heterostructures can easily introduce interfacial contamination[26], significantly degrading both the inherent and emergent properties.

In this study, we introduce a nano-confined chemical vapor deposition (CVD) approach that utilizes graphene or hBN as a vdW capping layer, enabling atomically-precise synthesis of 2D TMDs along with their simultaneous vdW integration. In contrast to conventional open growth, the nano-confined growth precisely yields $NbSe_2$ and $MoS_2$ monolayers owing to distinct kinetic mechanisms. The growth morphologies can be tailored from isolated monolayer domains to large-scale continuous films and intrinsically-patterned rings by simply tuning the growth conditions. Beyond classical TMDs, Janus MoSSe monolayers can be synthesized by selectively substituting the bottom chalcogen plane with the top plane protected by the vdW capping layer, demonstrating true atomic-plane selectivity. Importantly, the nano-confined growth simultaneously achieves vdW integration, creating graphene/TMD and hBN/TMD heterostructures with ultraclean vdW interfaces and enabling in situ encapsulation that effectively preserves air-sensitive materials. Benefiting from this reliable encapsulation, our nano-confined $NbSe_2$ monolayers exhibit exceptional air stability and a markedly higher superconducting transition temperature compared with conventional CVD-grown samples. Taken together, our results demonstrate that the nano-confined growth with vdW capping layers provides a versatile platform for the atomically-precise synthesis, ultraclean integration, and advanced applications of 2D TMDs.

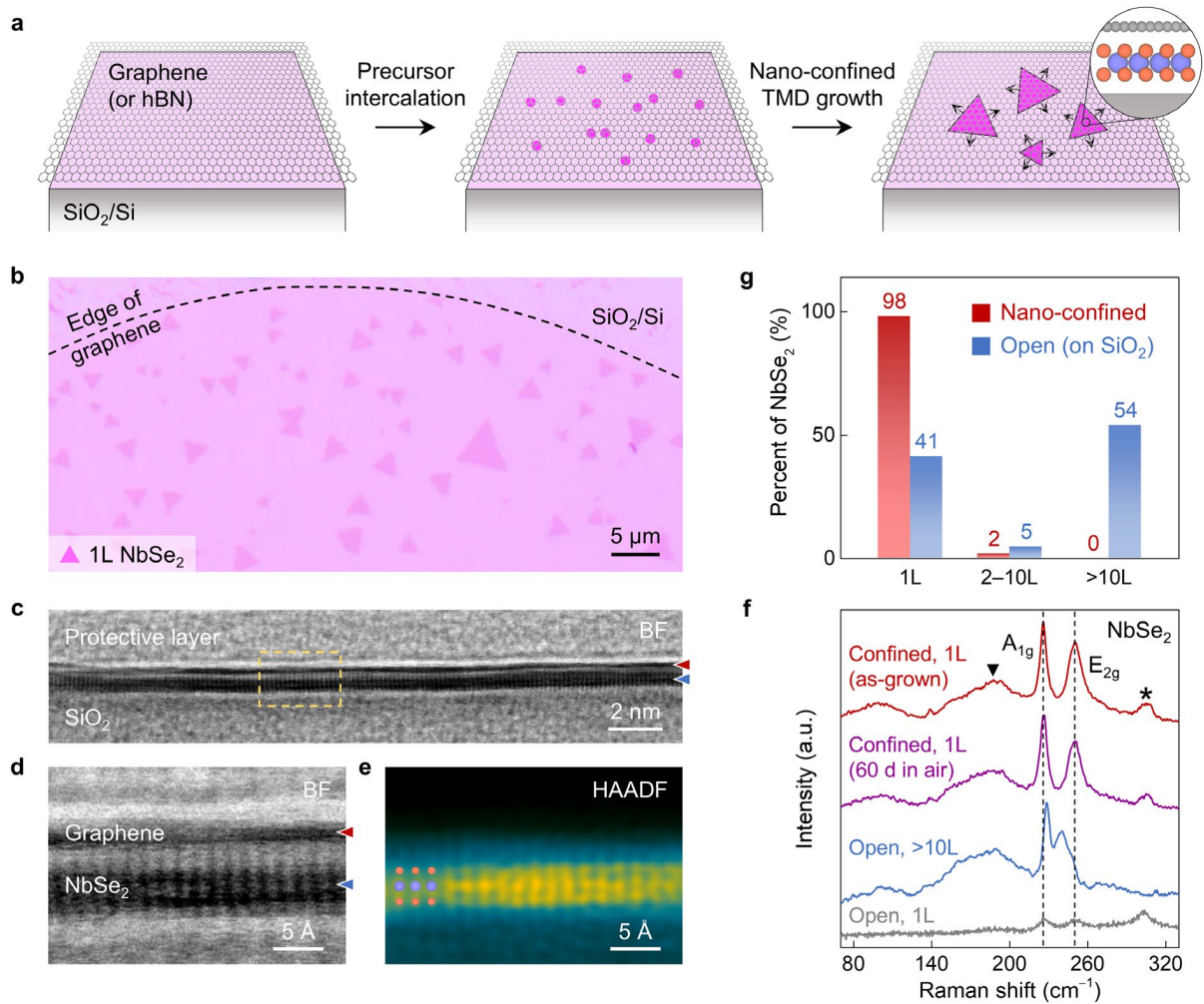

**Fig. 1 | Nano-confined growth of NbSe₂ monolayers. a**, Schematic diagram illustrating the growth procedure. **b**, Optical micrograph of NbSe₂ monolayers grown underneath a graphene monolayer. **c-e**, Cross-sectional STEM characterizations. **c**, Large-scale BF image. **d,e**, Close-up BF (**d**) and HAADF (**e**) images of the dashed region in **c**. The arrows in the BF images (**c,d**) indicate dark stripes from graphene and NbSe₂ monolayers. The atomic model of a NbSe₂ monolayer along the <210> zone axis is overlaid on the HAADF image (**e**), with the purple and orange balls representing the Nb and Se atoms, respectively. **f**, Raman spectra of nano-confined NbSe₂ monolayers and open-grown NbSe₂ crystals with different thicknesses. The dashed lines indicate the typical bands of NbSe₂ monolayers. ▼ and * indicate the NbSe₂ soft mode and the Si band, respectively. All spectra were collected under ambient conditions and right after growth unless otherwise specified. **g**, Thickness distributions of NbSe₂ crystals grown under the nano-confinement and on open SiO₂ substrates. $n$L denotes $n$ layer(s).

## Nano-confined growth of TMD monolayers

The nano-confined growth is schematically illustrated in Fig. 1a. First, graphene or hBN capping layers were mechanically exfoliated onto a SiO$_2$/Si substrate, creating an interface that served as the nano-confinement for TMD growth (left panel). Next, TMD precursors were produced by ambient-pressure CVD and intercalated into this interface (middle panel). As the precursor concentration increased, nucleation finally occurred, and monolayer TMDs formed with high precision under the confined environment (right panel).

As an example, the nano-confined growth of NbSe$_2$ monolayers is demonstrated underneath graphene capping layers. An optical micrograph of as-grown samples is shown in Fig. 1b, revealing triangular NbSe$_2$ domains with uniform optical contrast. The nano-confined growth scheme is confirmed by cross-sectional scanning transmission electron microscopy (STEM). In the bright-field (BF) STEM image (Fig. 1c,d), two dark stripes, indicated by arrows, are observed above the SiO$_2$ substrate. The simultaneously acquired high-angle annular dark-field (HAADF) image (Fig. 1e) clearly identifies the lower dark stripe as a NbSe$_2$ monolayer, with its atomic arrangement resolved along the <210> zone axis, and the upper dark stripe as the graphene capping layer, showing weak HAADF intensity consistent with a lower atomic number. This structural assignment is further verified by electron energy-loss spectroscopy (EELS) elemental mappings (Supplementary Fig. S6), collectively evidencing the formation of NbSe$_2$ monolayers underneath graphene.

The nano-confined NbSe$_2$ monolayers were characterized using Raman spectroscopy. Conventional open-grown NbSe$_2$ crystals with a broad thickness distribution (Supplementary Fig. S2)[27-29] were included for comparison. As shown in Fig. 1f, the nano-confined NbSe$_2$ monolayers exhibit the typical A$_{1g}$ and E$_{2g}$ bands at 226 and 250 cm$^{-1}$, respectively, along with the soft mode around 180 cm$^{-1}$ (red curve; Supplementary Fig. S4)[30]. The large A$_{1g}$–E$_{2g}$ band separation of 24 cm$^{-1}$ is characteristic of NbSe$_2$ monolayers, in contrast to open-grown thick NbSe$_2$ crystals, which show a small separation of 12 cm$^{-1}$ (blue curve in Fig. 1f)[30]. Importantly, atomically-thin NbSe$_2$ crystals are highly air-sensitive[31,32], as evidenced by the barely detectable Raman bands of open-grown NbSe$_2$ monolayers (gray curve in Fig. 1f). In contrast, the nano-confined NbSe$_2$ monolayers are in situ encapsulated and maintain sharp Raman bands even after 60 days of air exposure (purple curve in Fig. 1f; Supplementary Figs. S10 and S11). This exceptional air stability enables both extensive characterization and reliable device integration of the nano-confined NbSe$_2$ monolayers, markedly broadening their application potential.

The monolayer yields of different growth schemes are compared in Fig. 1g. Remarkably, 98% of nano-confined NbSe$_2$ crystals are monolayers, whereas open-grown counterparts exhibit a monolayer yield of only 41%. This contrast highlights distinct growth mechanisms under the nano-confinement, enabling precise monolayer formation regardless of macroscopic growth conditions.

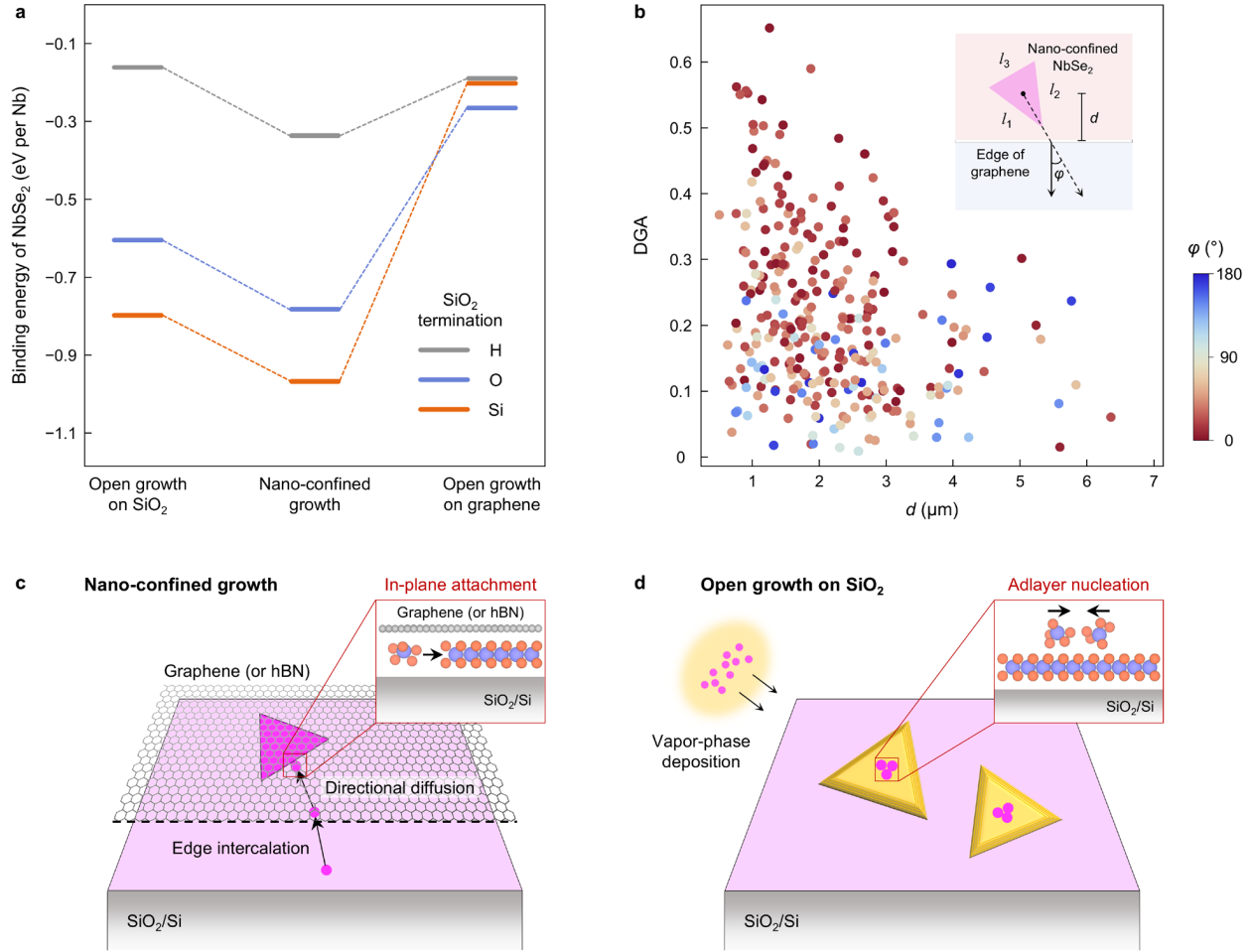

**Fig. 2 | Mechanisms of the nano-confined growth. a**, Comparison of the binding energy of NbSe$_2$ monolayers across different growth schemes, considering various SiO$_2$ terminations. **b**, Relationship between the degree of growth asymmetry (DGA) and the crystal-to-edge distance ($d$), with the color scale indicating the overgrowth direction ($\varphi$). Inset: definitions of $d$ and $\varphi$, as well as the side lengths ($l_1$, $l_2$, $l_3$) for determining DGA according to equation (1). The overgrowth direction (dashed arrow), pointing from the center of the crystal (black dot) to the overgrown vertex (intersection of $l_1$ and $l_2$), is measured relative to the graphene edge normal (solid arrow). **c,d**, Schematic diagrams illustrating growth kinetics for uniform TMD monolayers under the nano-confinement (**c**) and for multilayer TMD crystals on open SiO$_2$ substrates (**d**). The dashed line in **c** indicates the edge of the capping layer.

## Mechanisms of the nano-confined growth

The nano-confined growth mechanisms are analyzed from both energetic and kinetic perspectives. Energetically, the binding energy of NbSe$_2$ monolayers was calculated using density functional theory (DFT) for different growth schemes, including the nano-confined growth (graphene/NbSe$_2$/SiO$_2$), the open growth on SiO$_2$ (NbSe$_2$/SiO$_2$), and the open growth on graphene (NbSe$_2$/graphene/SiO$_2$), with various SiO$_2$ terminations to model the amorphous substrate (Supplementary Tables S2 and S3)[33]. As shown in Fig. 2a, the nano-confined NbSe$_2$ monolayers exhibit the most negative binding energy regardless of SiO$_2$ terminations (e.g., −0.97 eV/Nb for nano-confined growth vs. −0.80 eV/Nb for open growth on Si-SiO$_2$), demonstrating energetically favored growth under the nano-confinement, consistent with interactions at both interfaces of the TMD monolayers (Supplementary Fig. S13)[34].

From the kinetic perspective, the nano-confined growth relies on effective intercalation of precursors, which dictates the morphology of TMD crystals. Our study reveals two distinct morphologies: (1) triangular domains evenly distributed across graphene capping layers (Fig. 1b and Supplementary Fig. S18a) and (2) domains preferentially nucleated near graphene edges (Supplementary Fig. S2a). The former is attributed to surface intercalation of precursors facilitated by point defects or tears in the graphene basal plane, whereas the latter corresponds to intact graphene capping layers with the edges serving as the only intercalation paths. In this case, precursors diffuse directionally from the edges toward the interior, leading to asymmetric growth shapes of the nano-confined TMD crystals.

To confirm this edge-intercalation mechanism based on the symmetry of growth shapes, side lengths of triangular NbSe$_2$ monolayers were extracted from optical micrographs, and the degree of growth asymmetry (DGA) was quantified as follows:

$$\text{DGA} = \frac{\max_{i \neq j} |l_i - l_j|}{L}, \qquad (1)$$

where $l_i$ ($i$ = 1, 2, 3) are the side lengths and $L$ is their average value. The crystal-to-edge distance ($d$) and the overgrowth direction ($\varphi$) were extracted according to definitions shown in the inset of Fig. 2b. The relationship between DGA and $d$ is shown in the main panel of Fig. 2b, with $\varphi$ indicated by the color scale. Remarkably, highly asymmetric NbSe$_2$ crystals with DGA > 0.2 predominantly overgrow toward graphene edges (0°< $\varphi$ < 90°, red), highlighting the edges as effective intercalation paths and the directional precursor diffusion that promotes overgrowth at edge-facing vertices[35,36]. Consistently, our phase-field simulations incorporating edge intercalation reproduce this asymmetric growth behavior (Supplementary Fig. S14), further confirming the edge-intercalation mechanism.

Following edge intercalation and directional diffusion, the confined environment enables precursor incorporation exclusively through in-plane edge attachment (Fig. 2c), naturally preventing adlayer nucleation and ensuring precise monolayer formation. This contrasts with open CVD growth, where vaporized precursors are deposited directly onto the surface of TMD monolayers and readily form adlayers unless the deposition rate is stringently controlled (Fig. 2d and Supplementary Note S1)[36]. These results underscore the distinct growth kinetics under the nano-confinement, enabling precise synthesis of monolayer TMDs.

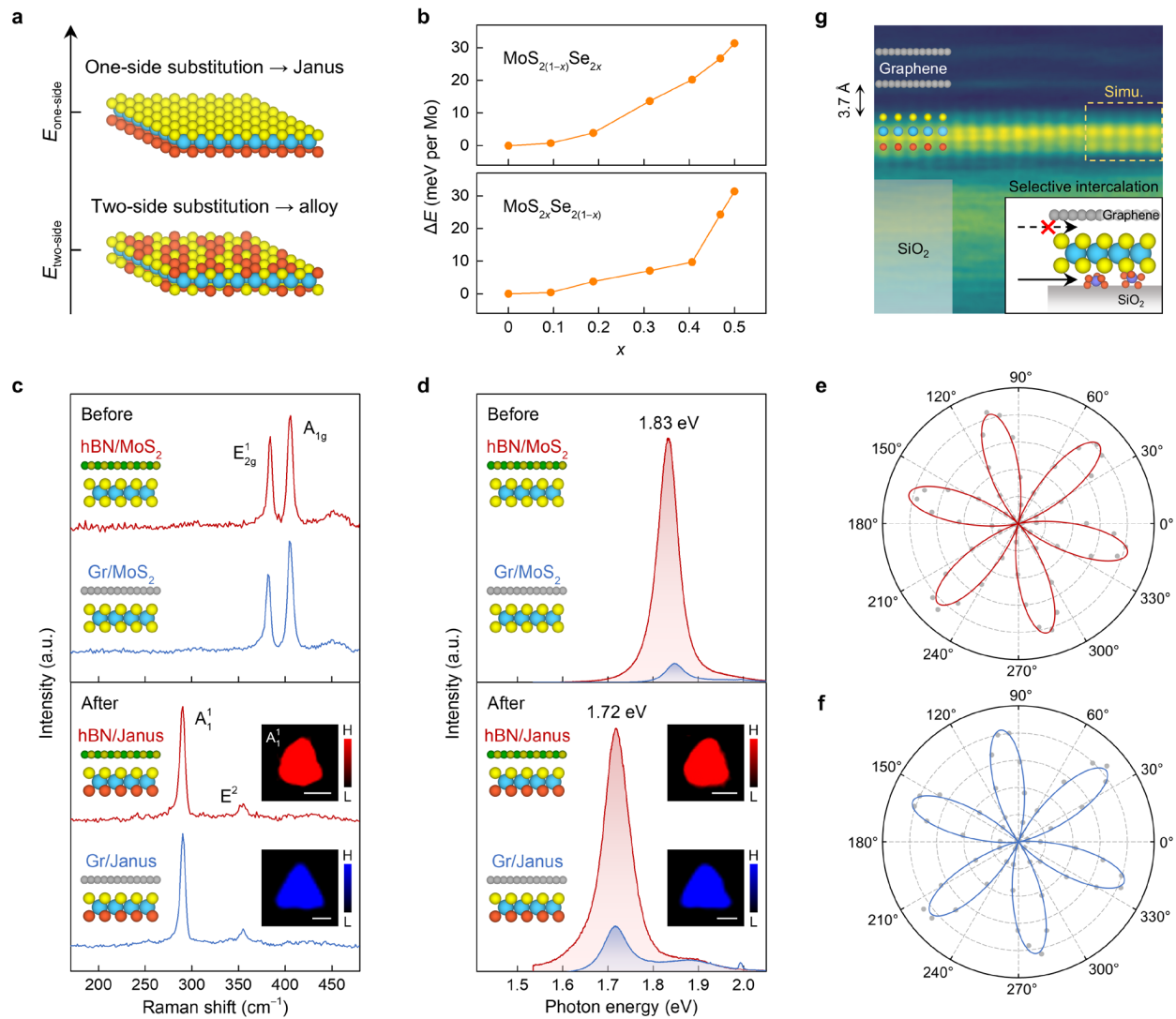

**Fig. 3 | Atomically-precise synthesis of Janus MoSSe monolayers under the nano-confinement. a**, Comparison of the Janus and alloy phases resulting from one-side and two-side substitution of a classical TMD monolayer. **b**, Substitution energy difference, defined as $\Delta E = E_{\text{one-side}} - E_{\text{two-side}}$, calculated for $MoS_{2(1-x)}Se_{2x}$ and $MoS_{2x}Se_{2(1-x)}$ (upper and lower panels, respectively). **c,d**, Raman (**c**) and PL (**d**) spectra of the as-synthesized $MoS_2$ monolayers (upper panels) and the Janus MoSSe monolayers after the substitution (lower panels), with either hBN or graphene as the capping layer. Left insets: sample schematics. Right insets: Raman and PL mappings. All scale bars are 1 μm. **e,f**, Polarization-resolved SHG intensities of the Janus MoSSe monolayers synthesized with hBN (**e**) or graphene (**f**) capping layers. The smooth curves represent fits to the data (Supplementary Fig. S16). **g**, Cross-sectional HAADF-STEM image of the graphene-confined Janus MoSSe monolayer, with a sample schematic overlaid on the left and a simulated HAADF image shown on the right. Bottom-right inset: schematic diagram illustrating the selective intercalation of precursors. The C, Mo, S, Nb, and Se atoms are represented by the gray, cyan, yellow, purple, and orange balls, respectively.

## Atomically-precise synthesis

Following the procedure shown in Fig. 1a, the nano-confined growth can be readily extended to other classical TMD monolayers (e.g., MoS$_2$; Supplementary Figs. S2, S4, S7, and S12) as well as to the use of CVD-grown graphene (Supplementary Fig. S3) and insulating hBN (Supplementary Figs. S2, S4, S8, S9, and S11) as the vdW capping layers, all governed by consistent energetic (Supplementary Tables S2-S5) and kinetic (Fig. 2c and Supplementary Figs. S2, S3 and S14) mechanisms. Moreover, the material library can be expanded to include Janus TMD monolayers featuring the polar chalcogen arrangement, for which the precision of the nano-confined growth reaches the atomic limit.

To date, the synthesis of Janus TMD monolayers has relied on the one-side substitution of classical TMD monolayers[15-20]. The challenge is that the Janus phase is not energetically favored compared to the alloy phase resulting from the random, two-side substitution (Fig. 3a). The energy difference between the one-side and two-side substitution products, defined as $\Delta E = E_{\text{one-side}} - E_{\text{two-side}}$, was calculated using DFT for different substitution ratios (Supplementary Tables S6 and S7). As shown in Fig. 3b, the upper (lower) panel presents $\Delta E$ for MoS$_{2(1-x)}$Se$_{2x}$ (MoS$_{2x}$Se$_{2(1-x)}$) monolayer, where $0 \leq x \leq 0.5$ is the substitution ratio with the lower and upper bounds denoting the pristine and half-substituted MoS$_2$ (MoSe$_2$) monolayer, respectively. It becomes evident that $\Delta E$ is always positive (i.e., $E_{\text{one-side}} > E_{\text{two-side}}$) and continuously increases with $x$, indicating that the one-side substitution of a classical TMD monolayer toward a Janus TMD monolayer is energetically unfavored, hence the key to realize the Janus phase is to control the substitution by kinetics rather than by energetics[15].

We demonstrate that the nano-confinement can be utilized to control the substitution kinetics, enabling the atomically-precise synthesis of Janus MoSSe monolayers. The process involves MoS$_2$ monolayers grown underneath graphene or hBN, with NbSe$_2$ precursors serving as substitution agents to supply Se atoms. Direct attachment of NbSe$_2$ precursors to the edges of MoS$_2$ monolayers is unlikely because of their substantial lattice mismatch (~9%)[37]. Instead, NbSe$_2$ precursors selectively intercalate into the MoS$_2$/SiO$_2$ interface, rather than into the vdW gap between graphene or hBN and MoS$_2$, owing to their distinct interfacial properties (Supplementary Note S3). As a result, only the bottom S plane is exposed to Se atoms, making the one-side substitution of the bottom S plane kinetically favored[17].

The as-synthesized Janus MoSSe monolayers were first characterized using Raman spectroscopy. As shown in Fig. 3c, the sharp $A_1^1$ Raman band at 290 cm$^{-1}$ arises from the out-of-plane vibrations of the S-Mo-Se bond (lower panel)[38], and the absence of MoS$_2$-related Raman bands (383 and 406 cm$^{-1}$; upper panel) demonstrates complete one-side substitution. Photoluminescence (PL) spectra before and after the substitution are compared in Fig. 3d. The emission peak shifts from 1.83 eV (MoS$_2$; upper panel) to 1.72 eV (Janus MoSSe; lower panel), agreeing with previous reports[19]. The notably narrower Raman and PL linewidths compared to literature values demonstrate the superior crystal quality of the as-synthesized Janus MoSSe monolayers (Supplementary Fig. S23). In addition, the superior substitution uniformity is confirmed by uniform Raman and PL intensities across the Janus MoSSe monolayers (see right insets of Fig. 3c,d for mappings). Because the one-side substitution preserves the in-plane crystal symmetry, similar 6-fold variations are observed in the polarization-resolved second-harmonic generation (SHG) intensity for both the Janus MoSSe monolayers (Fig. 3e,f and Supplementary Fig. S16) and the MoS$_2$ monolayers (Supplementary Fig. S12)[39]. Meanwhile, the out-of-plane crystal symmetry is broken by the one-side substitution, as directly revealed by cross-sectional HAADF-STEM. An atomically-resolved HAADF image of the graphene-confined Janus MoSSe monolayer is shown in Fig. 3g (see Supplementary Fig. S15 for the HAADF image of the hBN-confined sample). A close observation demonstrates that the chalcogen atoms on the two sides of the monolayer exhibit different HAADF intensities (Supplementary Fig. S15): the atoms on the lower side, showing stronger intensities, are identified as Se, whereas the upper atoms with weaker intensities are assigned to S. The corresponding HAADF simulation, overlaid on the dashed region in Fig. 3g, agrees with the experimental image. These results demonstrate the one-side substitution of the bottom S atoms, confirming the selective intercalation of NbSe$_2$ precursors into the MoS$_2$/SiO$_2$ interface (inset of Fig. 3g). Benefiting from the vdW capping layer, the top chalcogen atoms are effectively protected from unintentional modification, enabling the atomically-precise synthesis of high-quality Janus TMD monolayers, as evidenced by their narrow optical emission.

Moreover, the vdW gap between graphene and the Janus MoSSe monolayer was determined to be 3.7 Å (Fig. 3g), closely matching our DFT-calculated value of 3.4 Å (Supplementary Fig. S15), indicating an ultraclean vdW interface enabled by the simultaneous integration. This clean contact allows interfacial charge transfer, resulting in PL quenching in the graphene-confined Janus MoSSe monolayers, with similar behavior observed in the $MoS_2$/graphene heterostructures (Fig. 3d)[40]. This contrasts with the hBN-confined monolayers, which exhibit strongly enhanced PL emission owing to the insulating nature of hBN. Additionally, photoinduced doping governs the photoresponse of the graphene/Janus MoSSe heterostructure device, which provides further evidence of charge transfer and highlights its potential for emerging 2D optoelectronic applications (Supplementary Fig. S22).

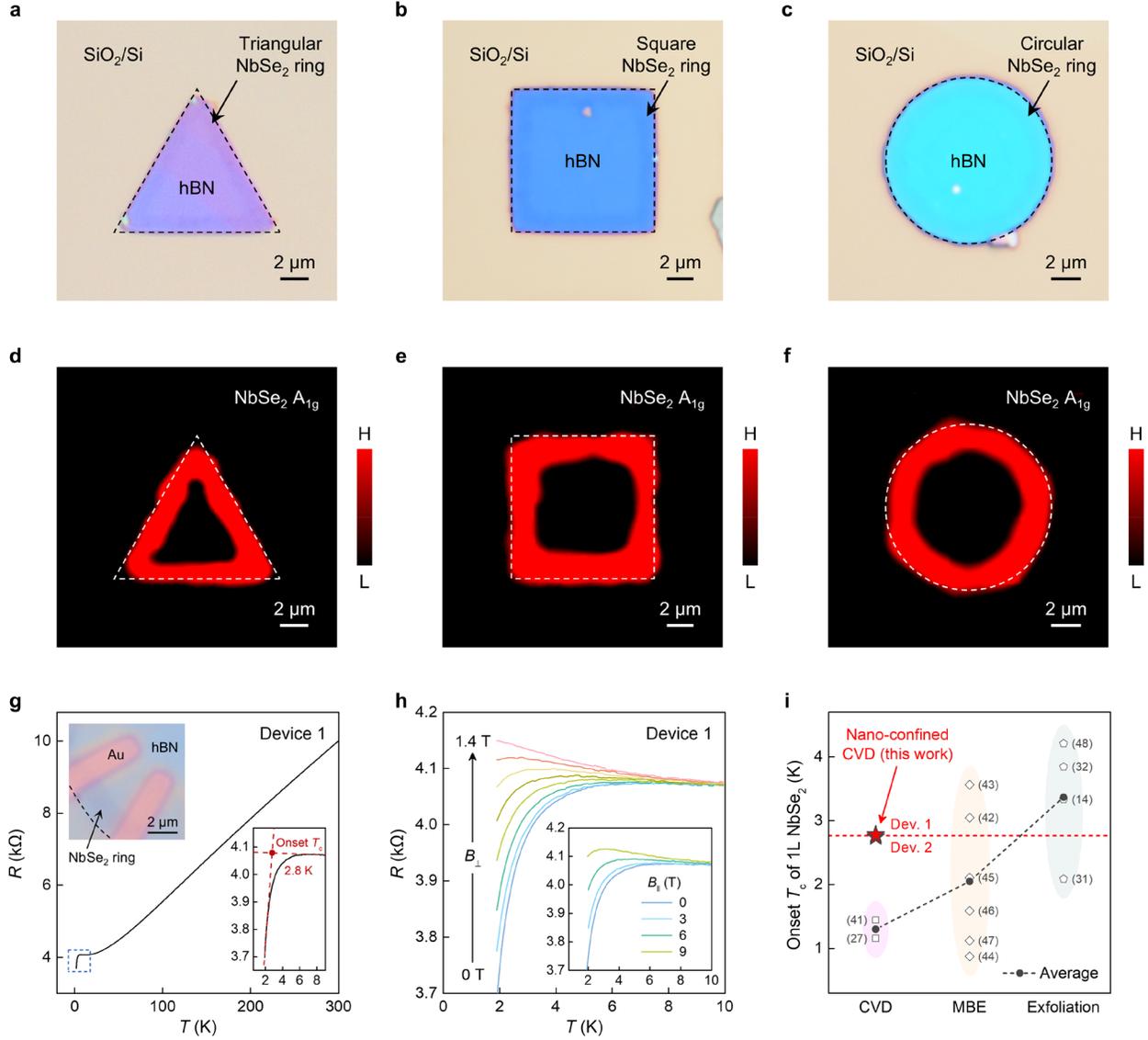

**Fig. 4 | Intrinsically-patterned growth of NbSe₂ monolayers. a-c**, Optical micrographs showing intrinsically-patterned NbSe₂ monolayer rings grown underneath hBN flakes with regular shapes. **d-f**, Raman mappings of NbSe₂ for the samples shown in **a-c**. Dashed lines in **a-f** indicate hBN edges. **g**, Resistance-temperature ($R$-$T$) curve of the intrinsically-patterned NbSe₂ ring. Bottom-right inset: close-up of the low-$T$ region (marked by dashed lines in the main panel), which highlights the onset of the superconducting transition. Top-left inset: optical micrograph of the device. The hBN edge is indicated by the dashed line. **h**, $R$-$T$ curves measured under perpendicular and parallel magnetic fields (main panel and inset, respectively). **i**, Comparison of onset $T_c$ values for NbSe₂ monolayers fabricated using different methods. Average values from this work and the literature are represented by red and black dashed lines, respectively. Device 2 data are provided in Supplementary Fig. S20.

## Intrinsically-patterned growth

By increasing the deposition flux of precursors to promote coalescence of isolated TMD domains, the nano-confined growth can be extended to two distinct growth regimes: (1) large-scale synthesis of continuous TMD films under defective capping layers that facilitate homogeneous nucleation (Supplementary Fig. S18) and (2) intrinsically-patterned growth of TMD rings under intact capping layers that direct nucleation near their edges (Supplementary Fig. S19). These growth morphologies are captured in a kinetic phase diagram defined by the defect density in the capping layers and the precursor deposition flux (Supplementary Fig. S17). Notably, the intrinsic-patterning regime enables the precise synthesis of patterned TMD monolayers simply by designing the capping layer geometry, avoiding post-synthesis lithography or etching processes that often degrade the monolayer quality.

Figure 4a-c shows optical micrographs of intrinsically-patterned $NbSe_2$ monolayers, which clearly follow the edge contours of the hBN capping layers with regular geometric shapes. Corresponding Raman mappings shown in Fig. 4d-f demonstrate uniform intensity across the $NbSe_2$ rings, which confirms their structural continuity and enables their application as conductive channels in 2D circuits. As a proof-of-concept, devices were fabricated based on the intrinsically-patterned $NbSe_2$ rings by transferring the hBN/$NbSe_2$ heterostructures onto bottom metal electrodes. An optical micrograph of a typical device is shown in the top-left inset of Fig. 4g. The main panel presents the temperature-dependent resistance, which reveals a linear relationship consistent with the metallic nature of $NbSe_2$, followed by a sharp superconducting transition with an onset transition temperature ($T_c$) of 2.8 K (bottom-right inset of Fig. 4g). Under perpendicular magnetic fields, the superconducting transition is quickly suppressed (main panel of Fig. 4h), whereas it is substantially more robust under parallel fields (inset of Fig. 4h) owing to robust Ising pairing protected by spin-orbit interaction[14]. To assess the superconducting performance of our $NbSe_2$ monolayers, the onset $T_c$ was benchmarked against that of samples fabricated using alternative methods. As shown in Fig. 4i, the onset $T_c$ of our $NbSe_2$ monolayers markedly exceeds those synthesized using conventional CVD techniques[27,41], matches molecular-beam epitaxy (MBE)-grown samples[42-47], and even approaches mechanically exfoliated $NbSe_2$ monolayers[14,31,32,48]. Such excellent performance highlights the enhanced uniformity, superior

quality, and exceptional stability of the nano-confined NbSe$_2$ monolayers (Supplementary Table S9), paving the way for their applications in superconducting circuits.

## Conclusions

We establish a nano-confined approach to synthesizing TMD monolayers using graphene or hBN as the vdW capping layer. The growth mechanisms are governed by distinct kinetics—including edge intercalation, directional diffusion, and in-plane attachment of precursors—which naturally ensure the precise formation of monolayers, achieving a high yield of 98% for NbSe$_2$. The realization of Janus MoSSe monolayers through vdW-protected substitution on a single chalcogen plane of MoS$_2$ highlights the atomic-level precision of the nano-confined synthesis. The vdW capping layers also serve as in situ encapsulation, effectively preserving the superior crystal quality of the air-sensitive NbSe$_2$ monolayers. This enables exceptional air stability for over 60 days after synthesis and allows the observation of enhanced superconducting performance, with an onset $T_c$ of 2.8 K that markedly exceeds reported values for CVD-grown NbSe$_2$ monolayers. Altogether, the nano-confined growth offers a novel platform for the atomically-precise synthesis and ultraclean integration of monolayer TMDs (Supplementary Table S8). We note that wafer-scale CVD growth of graphene and hBN has been well established[49,50], and the resulting films can be transferred onto SiO$_2$/Si substrates with high cleanliness and uniformity[51,52]. Integrating these mature technologies for the capping layers with the nano-confined growth should enable wafer-scale synthesis of continuous TMD films when basal-plane defects are controllably introduced into the capping layers[53,54], as well as the scalable fabrication of TMD-based 2D circuits by leveraging the unique intrinsic-patterning capability. This work highlights the nano-confined growth of compound TMD monolayers, which, together with the confined growth of elemental 2D materials (e.g., Si, Ge, Pt, Pd, Ni, Co, Au, In, and Ce) at graphene/metal interfaces[55-57], establishes a promising platform for engineering novel quantum structures and enabling functional quantum devices.

## Methods

**Preparation of capping layers.** Graphene and hBN flakes were mechanically exfoliated onto 285-nm SiO$_2$/Si substrates using the Scotch tape method. The as-exfoliated samples were annealed under vacuum (<5×10$^{-1}$ Pa) at 500 °C for 1 h to remove tape residue.

**Nano-confined growth of NbSe$_2$.** The nano-confined growth of NbSe$_2$ monolayers was conducted using an ambient-pressure CVD system equipped with a two-zone tube furnace (Supplementary Fig. S1 and Table S1). The Nb source, a well-ground powder of Nb$_2$O$_5$, Nb, and NaCl (weight ratio of 2:2:1; ~2 mg), was placed in a quartz boat, and the graphene- or hBN-covered SiO$_2$/Si substrate was placed above the powder with the polished surface facing down. This quartz boat was loaded into the downstream zone. Se powder (~0.1 g) in another quartz boat was loaded into the upstream zone. After purging the quartz tube with 10% H$_2$/Ar for 10 min, the up- and downstream zones were heated to 220 and 805 °C in 15 min, respectively, and held for 3~5 min before being cooled down to room temperature naturally. 10% H$_2$/Ar with a flow rate of 150 standard cubic centimeters per minute (SCCM) was used as the carrier gas. The open-grown NbSe$_2$ crystals were obtained on bare regions of the SiO$_2$/Si substrate without being covered by graphene or hBN flakes.

**Nano-confined growth of MoS$_2$.** The nano-confined growth of MoS$_2$ monolayers was conducted using a similar CVD system (Supplementary Fig. S1 and Table S1). A thin film of MoO$_3$ grown on mica was used as the Mo source. The MoO$_3$ film was placed in a quartz boat, and the SiO$_2$/Si substrate was placed above the MoO$_3$ film. This quartz boat was loaded into the downstream zone, and another quartz boat containing S powder (~0.2 g) was loaded into the upstream zone. The quartz tube was then purged with Ar for 10 min after which the up- and downstream zones were heated to 140 and 850 °C in 20 min, respectively, and held for 10~15 min before they were naturally cooled down to room temperature. Ar (100 SCCM) was used as the carrier gas.

**Nano-confined growth of Janus MoSSe.** To synthesize Janus MoSSe monolayers, an as-grown MoS$_2$ sample was loaded into the CVD system of NbSe$_2$ and placed in the downstream zone (Supplementary

Fig. S1 and Table S1). The downstream-zone temperature was reduced to 750 °C to suppress NbSe$_2$ growth, while still allowing the substitution. All other steps and parameters remained unchanged.

**Intrinsically-patterned growth.** Intrinsically-patterned growth requires both intact capping layers and high deposition flux. The integrity of the capping layers was ensured using multilayer flakes, which effectively prevent point defects and tears from penetrating through the basal plane, and the high deposition flux was achieved by increasing the amount of the transition metal source. For the intrinsically-patterned growth of NbSe$_2$ underneath hBN, multilayer hBN flakes were predefined into regular geometric shapes using photolithography (Karl Suss MA6/BA6) followed by CHF$_3$/O$_2$ reactive-ion etching (RIE; Oxford Instruments Plasma Lab 80Plus). The resulting hBN flakes were placed above the Nb source with the amount increased to ~20 mg to achieve the high flux condition (Supplementary Table S1). All other steps and parameters followed those described for NbSe$_2$ growth.

**Sample characterizations.** Optical micrographs were obtained using a 100× objective lens of an Olympus BX51 microscope equipped with a halogen light source and an SC30 color camera. The optical contrast of crystals extracted from the optical micrographs was used to determine the crystal thickness (Supplementary Fig. S5). To analyze the DGA of the nano-confined crystals, $l_1$, $l_2$, $l_3$, as well as $\varphi$ and $d$ were determined from the optical micrographs. Raman and PL spectroscopies were carried out under ambient conditions using a commercial instrument (WITec alpha300 R) equipped with a 532-nm excitation laser. Polarization-resolved SHG characterization was conducted using a home-built setup (Supplementary Fig. S12) equipped with a 1064-nm excitation laser (NPI Rainbow 1064 OEM). Cross-sectional samples for transmission electron microscopy were prepared with the lift-out technique using a focused ion beam (FIB; Helios G4 CX DualBeam). Aberration-corrected STEM imaging and EELS mapping were performed using a Nion Ultra-HERMES-100 microscope operated at 100 kV. HAADF-STEM images were acquired using a converge angle of 30 mrad and an annular detector with 92~210 collection semi-angles. The software STEM_CELL[58] was used to perform the HAADF simulation of Janus MoSSe monolayer according to the experimental settings of the microscope. The sample thickness was set to 10 nm. The thermal diffuse scattering was set to 10 cycles. The source size broadening was considered in the simulation by including a Gaussian spread with a full width at half maximum (FWHM) of 0.8 Å.

**Theoretical calculations.** DFT calculations were performed using the Vienna Ab initio Simulation Package (VASP)[59] and the projector augmented-wave (PAW) method[60]. The exchange-correlation potential was described using the Perdew-Burke-Ernzerhof (PBE) form of the generalized gradient approximation (GGA)[61]. The vdW correction was considered using the DFT-D2 approach[62]. For the calculation of the binding energy of NbSe$_2$ monolayer, an energy cutoff of 400 eV was adopted for the plane-wave basis set, and the Brillouin zone was sampled using a 7×7×1 $k$-point mesh. The energy and force convergence thresholds were set to $1.0\times10^{-4}$ eV and 0.01 eV/Å, respectively. The binding energy was obtained according to $E_{binding} = E_{total} - E_{TMD} - \sum E_{neighbor}$, where $E_{total}$ is the total energy of the adsorbing system, $E_{TMD}$ is the energy of the isolated TMD monolayer, and $E_{neighbor}$ is the energy of the isolated neighboring layer. As for the calculations of the energy of the substituted TMD monolayers (i.e., MoS$_{2(1-x)}$Se$_{2x}$ and MoS$_{2x}$Se$_{2(1-x)}$) and the graphene/Janus MoSSe vdW gap, an energy cutoff of 300 eV was adopted, and the Brillouin zones were sampled using 2×2×1 and 3×3×1 $k$-point meshes, respectively. The energy convergence threshold was set to $1.0\times10^{-6}$ eV, and the force convergence threshold was set to 0.01 eV/Å. The vacuum layer thickness was >10 Å for all DFT calculations. Phase-field simulations of the nano-confined growth shape and the precursor distribution were performed using a two-region model (Supplementary Fig. S14) that incorporates the edge-intercalation process. The detailed simulation method can be found in the previous work[63].

**Device fabrication and measurement.** For the graphene/Janus MoSSe heterostructure devices, fabrication and measurement procedures are detailed in Supplementary Fig. S21. The intrinsically-patterned NbSe$_2$ devices were fabricated by transferring the as-grown hBN/NbSe$_2$ heterostructures onto bottom metal electrodes using a polycarbonate (PC)-based stamping method[64]. The electrodes were prepared using photolithography (Karl Suss MA6/BA6) and electron-beam evaporation (AST Peva-600E; Cr/Au 5/20 nm), followed by a lift-off process in acetone. Device measurements were conducted using the Quantum Design Physical Property Measurement System (PPMS) with the lowest accessible temperature of ~2 K. Resistance was measured in the two-terminal configuration using the Resistivity Option of the PPMS in AC mode, with a typical excitation current of 100 nA.

## Acknowledgements

We thank Prof. Min Ouyang for helpful discussions. This work was supported by the National Key R&D Program of China (2022YFA1204100), the National Natural Science Foundation of China (62488201), and the Chinese Academy of Sciences (XDB33030100 and XDB30010000). Y.Z. and F.D. acknowledge the startup grant and the High-Talent Grant (SIAT-SE3G0991010,2023) from the Shenzhen Institute of Advanced Technology. R.G. and W.Z. acknowledge support from the National Natural Science Foundation of China (52373231) and the Beijing Outstanding Young Scientist Program (BJJWZYJH01201914430039). H.H. acknowledges support from the Chinese Academy of Sciences (XDB30000000). W.H. and Z.X. acknowledge support from the National Natural Science Foundation of China (52090032). This work was benefited from support and resources from the Electron Microscopy Center at the University of Chinese Academy of Sciences.

## Author contributions

H.-J.G. conceived the idea and designed the project. C.B., Q.Q., K.Z., H.W., K.W., and P.P. synthesized the samples. C.B. performed the Raman, PL, and SHG characterizaions and conducted the crystal thickness and DGA analyses. R.G., Z.W., and W.Z. performed the STEM/EELS characterizations. Y.Z., H.H., and F.D. performed the DFT calculations. W.H. and Z.X. performed the phase-filed simulaitons. C.B. and H.L. fabricated and measured the devices. C.B., H.L., H.G., F.D., and H.Y. analyzed the growth mechanisms. C.B., H.L., W.Z., H.Y., and H.-J.G. wrote the paper with input from all authors.

## Competing interests

The Institute of Physics, Chinese Academy of Sciences filed a Chinese patent application (202111202207.3), which lists C.B., H.Y., and H.-J.G. as the inventors. Other than that, the authors declare no competing interests.

## Additional information

**Correspondence and requests for materials** should be addressed to Wu Zhou, Feng Ding, Haitao Yang, or Hong-Jun Gao.

# Supplementary information
# Atomically-precise synthesis and simultaneous integration of 2D transition metal dichalcogenides enabled by nano-confinement

Ce Bian[1,2,#], Yifan Zhao[3,4,#], Roger Guzman[2,#], Hongtao Liu[1,#], Hao Hu[5], Qi Qi[1,2], Ke Zhu[1,2], Hao Wang[1,2], Kang Wu[1,2], Hui Guo[1], Wanzhen He[6], Zhaoqing Wang[2], Peng Peng[3,7], Zhiping Xu[6], Wu Zhou[2,*], Feng Ding[3,4,7*], Haitao Yang[1,2,*] and Hong-Jun Gao[1,2,*]

[1] *Beijing National Center for Condensed Matter Physics and Institute of Physics, Chinese Academy of Sciences, Beijing, China*

[2] *School of Physical Sciences, University of Chinese Academy of Sciences, Beijing, China*

[3] *Suzhou Laboratory, Suzhou, China*

[4] *Institute of Technology for Carbon Neutrality, Shenzhen Institute of Advanced Technology, Chinese Academy of Sciences, Shenzhen, China*

[5] *Frontier Institute of Science and Technology, Xi'an Jiaotong University, Xi'an, China*

[6] *Department of Engineering Mechanics and Center for Nano and Micro Mechanics, Tsinghua University, Beijing, China*

[7] *Faculty of Materials Science and Energy Engineering, Shenzhen University of Advanced Technology, Shenzhen, China*

[#]These authors contributed equally: Ce Bian, Yifan Zhao, Roger Guzman, Hongtao Liu

[*]E-mail: wuzhou@ucas.ac.cn; dingf@szlab.ac.cn; htyang@iphy.ac.cn; hjgao@iphy.ac.cn

# Table of contents

# Part I | Background

**Note S1 | CVD synthesis of TMDs with the precise monolayer thickness**

Substantial efforts have been devoted to the precise synthesis of 2D TMDs. Of particular interest is the precise synthesis of TMD monolayers using the CVD technique, which has been demonstrated to be a cost-effective, high-yield, and versatile approach for the "bottom-up" fabrication of various nano-materials[1]. During the CVD growth of TMDs, vaporized precursors can be deposited onto the substrate as well as the TMD monolayers. A significant amount of precursors will be present on the TMD monolayers if the rate of deposition is high enough. Under this condition, the precursors can easily arrive at a same position to form a cluster, which may result in the growth of an adlayer if the size of the cluster exceeds the critical size for nucleation[2]. Several approaches have been developed to reduce the deposition rate of precursors and thereby suppress adlayer nucleation, such as (1) the use of a low chalcogen source temperature which limits the deposition rate of chalcogen vapor[3], (2) the use of sophisticated metal-organic CVD (MOCVD) wherein the deposition rate can be reduced by lowering the partial pressure of precursors[4] and (3) the use of a spatially-confined reactor (e.g., a ~0.5-mm gap between a mica substrate and a quartz tube) which provides a local environment with a lower precursor concentration and hence a slower deposition rate[5,6]. A different approach is to use the pulsed deposition procedure in which uniform TMD monolayers can be synthesized using high deposition rates[7]. Between pulses is the "relaxation" period with no precursors supplied. During this period, the precursors on the surface of a TMD monolayer can diffuse sufficiently and attach to the edges of the monolayer, the energetically-favored growth sites, to contribute to the in-plane growth. Regardless of the approach adopted, the competition between the deposition of precursors and their subsequent surface diffusion toward the edges dictates the rate of adlayer nucleation[2]. The precise synthesis of monolayer TMDs can be achieved only when the latter is faster than the former. An unusual growth mechanism, by which the precursors can attach directly to the edges without being deposited onto the surface, can naturally lead to the precise synthesis of TMD monolayers free from the difficulties caused by the fine tuning of growth parameters and the involvement of complicated equipment. Such unique growth kinetics is highly desired, but it remains unreported in the literature.

**Note S2 | Atomically-precise synthesis of Janus TMD monolayers**

The Janus TMD monolayers can be further synthesized through precise substitution of the classical TMD monolayers, a delicate process requiring complete replacement of a single plane of chalcogen atoms while preserving the opposite plane intact. To achieve such atomically-precise substitution, the substitution agents (i.e., precursors with the second chalcogen species) have to be supplied to only one side of the TMD monolayer, such that, under appropriate conditions, only the chalcogen atoms in direct contact with the precursors can be substituted owing to a lower kinetic barrier compared to that of the chalcogen exchange across the transition metal plane[8]. Based on this approach, different pathways for precursor supply have been developed to synthesize Janus TMD monolayers, such as (1) supplying the precursors via the vapor phase, either chalcogen vapor[9,10], chalcogen plasma[8], or a mixture of chalcogen vapor and $H_2$ plasma[11,12], meanwhile the substrate (e.g., $SiO_2$ and $Al_2O_3$) is utilized as a mask and, alternatively, (2) supplying the precursors through the TMD/Au interface in which the Au substrate catalytically dissociates the chalcogen clusters and facilitates the migration of chalcogen atoms[13]. The substitution is intended to take place on the top and bottom sides of the TMD monolayer for (1) and (2), respectively; however, it is still unclear to what extent the opposite side is kept intact during the substitution. For (1), the bottom chalcogen atoms will be modified if the precursors intercalate into the TMD/oxide interface[10]. For (2), the undissociated chalcogen clusters, which remain chemically active at elevated temperatures[9,10], are deposited directly onto the surfaces of TMD monolayers—potentially resulting in modification of the top chalcogen atoms. The likely presence of undesired processes in the previous works indicates partial control over the substitution kinetics, which is insufficient to achieve the atomically-precise synthesis of Janus TMD monolayers.

# Part II | Additional data and discussions

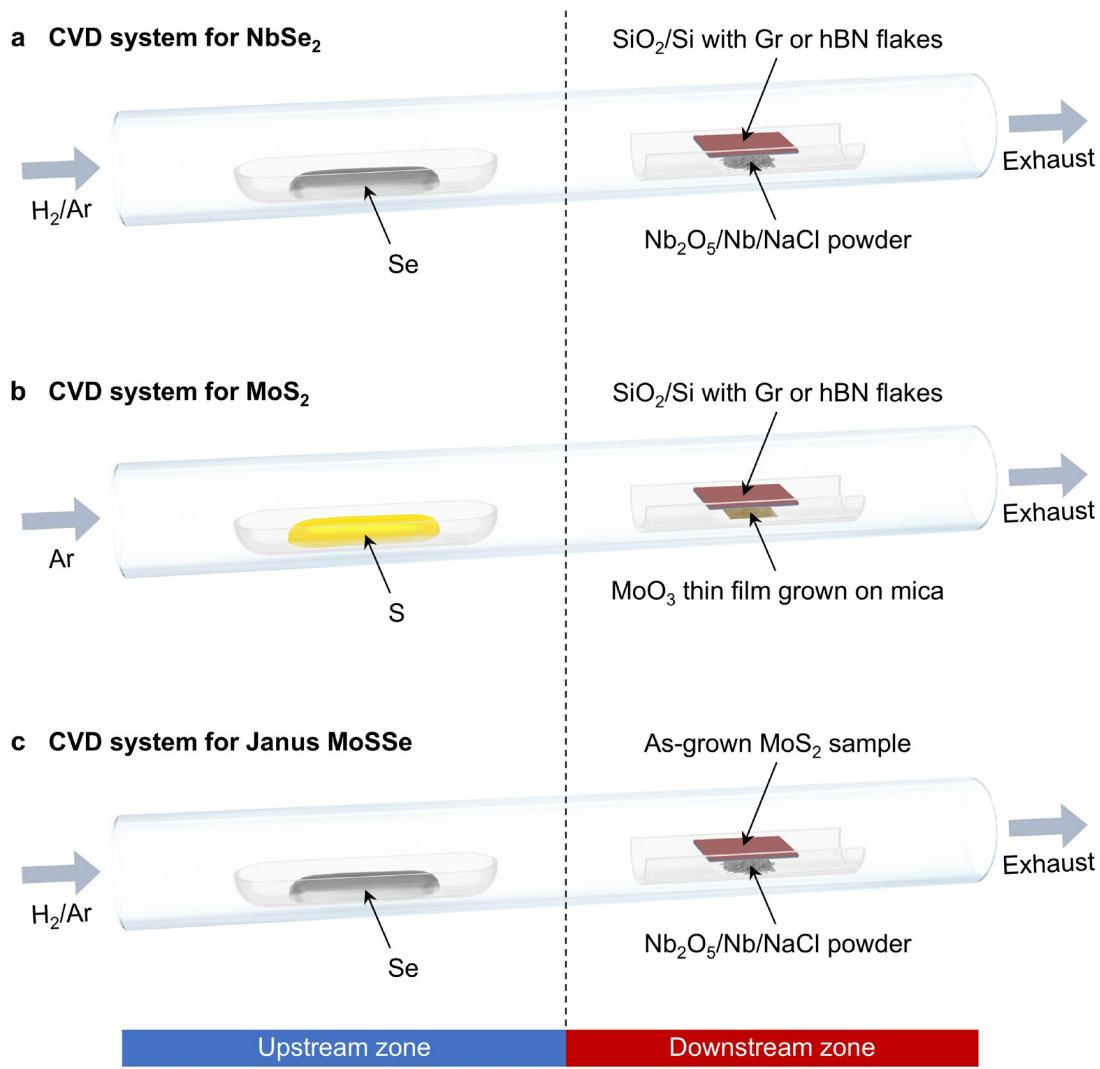

**Fig. S1 | CVD systems.** Schematic diagrams showing the CVD systems used for the nano-confined growth of NbSe$_2$ (**a**), MoS$_2$ (**b**), and Janus MoSSe (**c**) monolayers. For the growth of NbSe$_2$, the Nb source was a well-ground powder of Nb$_2$O$_5$, Nb, and NaCl. The adding of NaCl could facilitate the evaporation of the Nb source[3,14]. As for the growth of MoS$_2$, the Mo source was a MoO$_3$ thin film grown on a mica substrate. The detailed preparation procedure can be found in the previous work[15].

**Table S1 | Growth parameters.** Growth parameters used for the nano-confined growth of $NbSe_2$, $MoS_2$, and Janus MoSSe monolayers.

| TMD | Substrate | Chalcogen source | Transition metal source | Carries gas | Temperature (Up-/downstream) | Holding time |
|---|---|---|---|---|---|---|
| $NbSe_2$ | $SiO_2$/Si with Gr or hBN flakes | Se powder ~0.1 g | $Nb_2O_5$/Nb/NaCl (2:2:1) powder 2−20 mg | 10% $H_2$/Ar 150 SCCM | 220/805 °C | ~3 min |
| $MoS_2$ | $SiO_2$/Si with Gr or hBN flakes | S powder ~0.2 g | $MoO_3$ thin film grown on mica | Ar 100 SCCM | 140/850 °C | ~10 min |
| Janus MoSSe | As-grown $MoS_2$ sample | Se powder ~0.1 g | $Nb_2O_5$/Nb/NaCl (2:2:1) powder ~2 mg | 10% $H_2$/Ar 150 SCCM | 220/750 °C* | ~3 min |

* The downstream-zone temperature was reduced to suppress $NbSe_2$ growth, while still allowing the substitution.

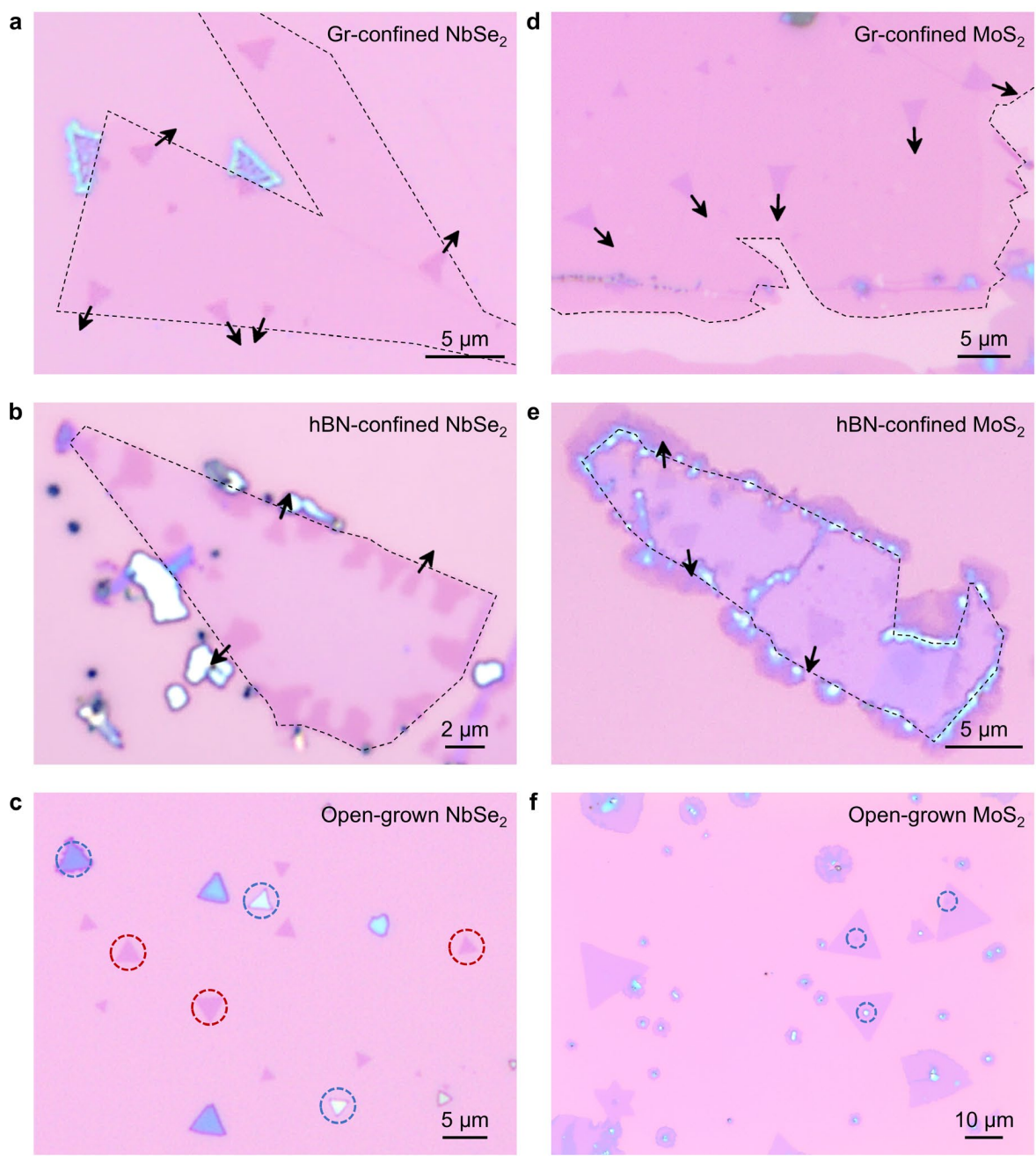

**Fig. S2 | Optical micrographs comparing different growth schemes. a-c**, Nano-confined growth (**a,b**) and open growth (**c**) of NbSe₂ using identical parameters. The NbSe₂ crystals grown under the nano-confinement are precisely monolayer despite the use of different capping layers (graphene in **a** and hBN in **b**), whereas with open SiO₂ substrates, multilayer and thick NbSe₂ crystals (blue circles in **c**) can be grown along with the monolayers (red circles in **c**). Owing to the edge intercalation of precursors and their directional diffusion underneath graphene and hBN, the nano-confined NbSe₂

monolayers, especially those close to the graphene and hBN edges (dashed lines in **a** and **b**), exhibit asymmetric triangular growth shapes with the directions of overgrowth (arrows in **a** and **b**) toward the edges. On the contrary, the open-grown NbSe$_2$ crystals have nearly-symmetric growth shapes which imply a more uniform distribution of precursors outside the confinement. **d-f**, Nano-confined growth (**d**,**e**) and open growth (**f**) of MoS$_2$ using identical parameters. Similarly, the distinct growth mechanisms under the nano-confinement lead to the precise synthesis of MoS$_2$ monolayers as well as the overgrowths (arrows in **d** and **e**), while the undesired growth of adlayers (blue circles in **f**) can easily take place in the open environment which considerably affects the monolayer uniformity.

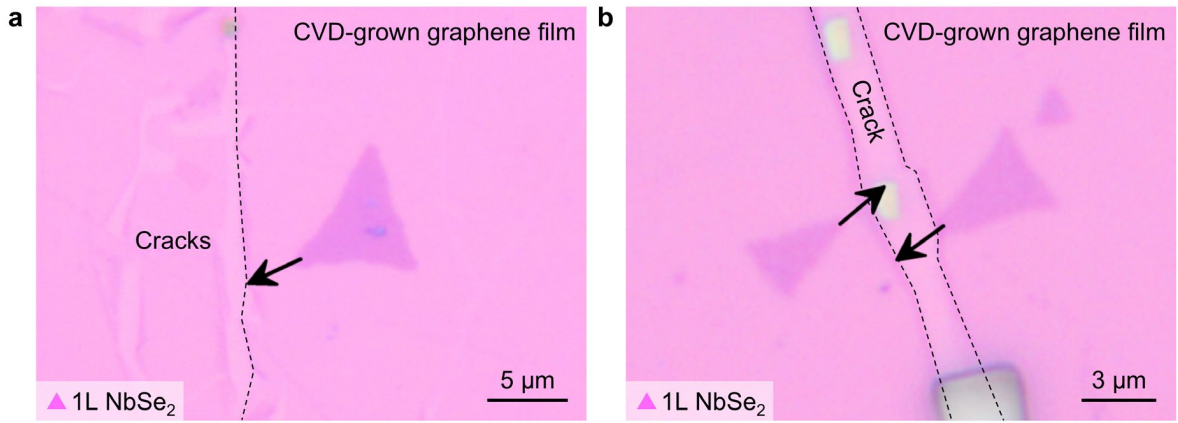

**Fig. S3 | Extension to CVD-grown capping layers. a,b**, Optical micrographs of NbSe$_2$ monolayers grown underneath CVD-grown graphene films, transferred from Cu foil onto SiO$_2$/Si substrates using a standard wet-transfer method[16]. The NbSe$_2$ monolayers preferentially nucleate near graphene edges and exhibit asymmetric shapes, similar to the growth behavior observed under exfoliated graphene (Figs. 2b and S2), confirming that the nano-confined growth is governed by the same kinetic mechanism (Fig. 2c). The overgrowth direction is indicated by the arrows. Dashed lines highlight graphene edges.

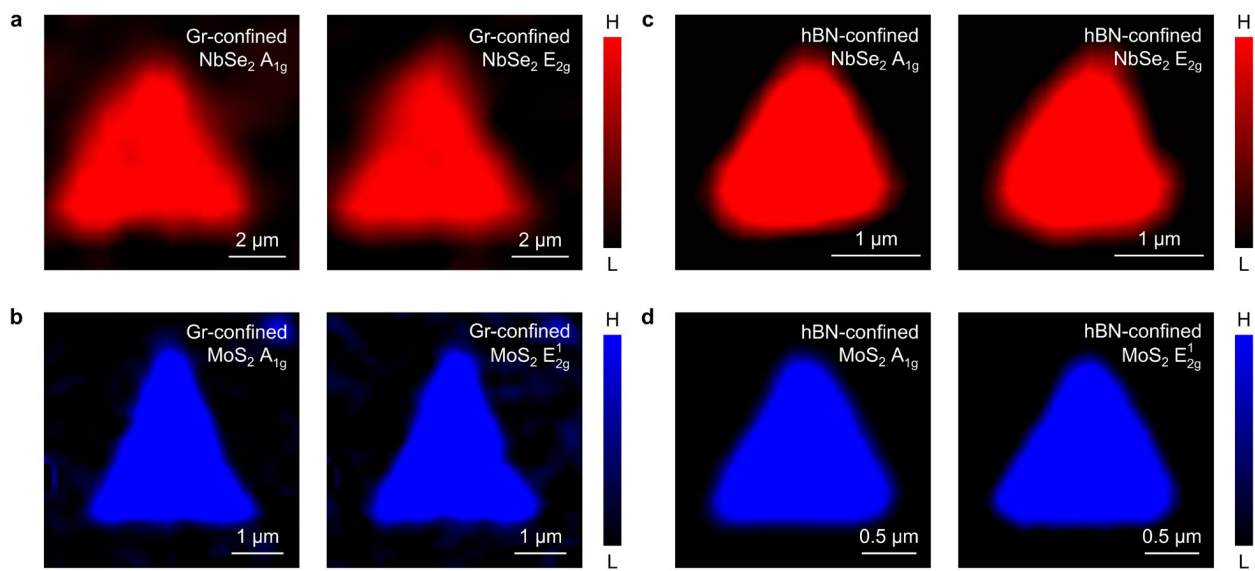

**Fig. S4 | Raman mappings of nano-confined NbSe$_2$ and MoS$_2$. a,b**, Graphene-confined NbSe$_2$ (**a**) and MoS$_2$ (**b**) monolayers. **c,d**, hBN-confined NbSe$_2$ (**c**) and MoS$_2$ (**d**) monolayers. Measurements were carried out under ambient conditions. Each Raman mapping shows spatially uniform intensity.

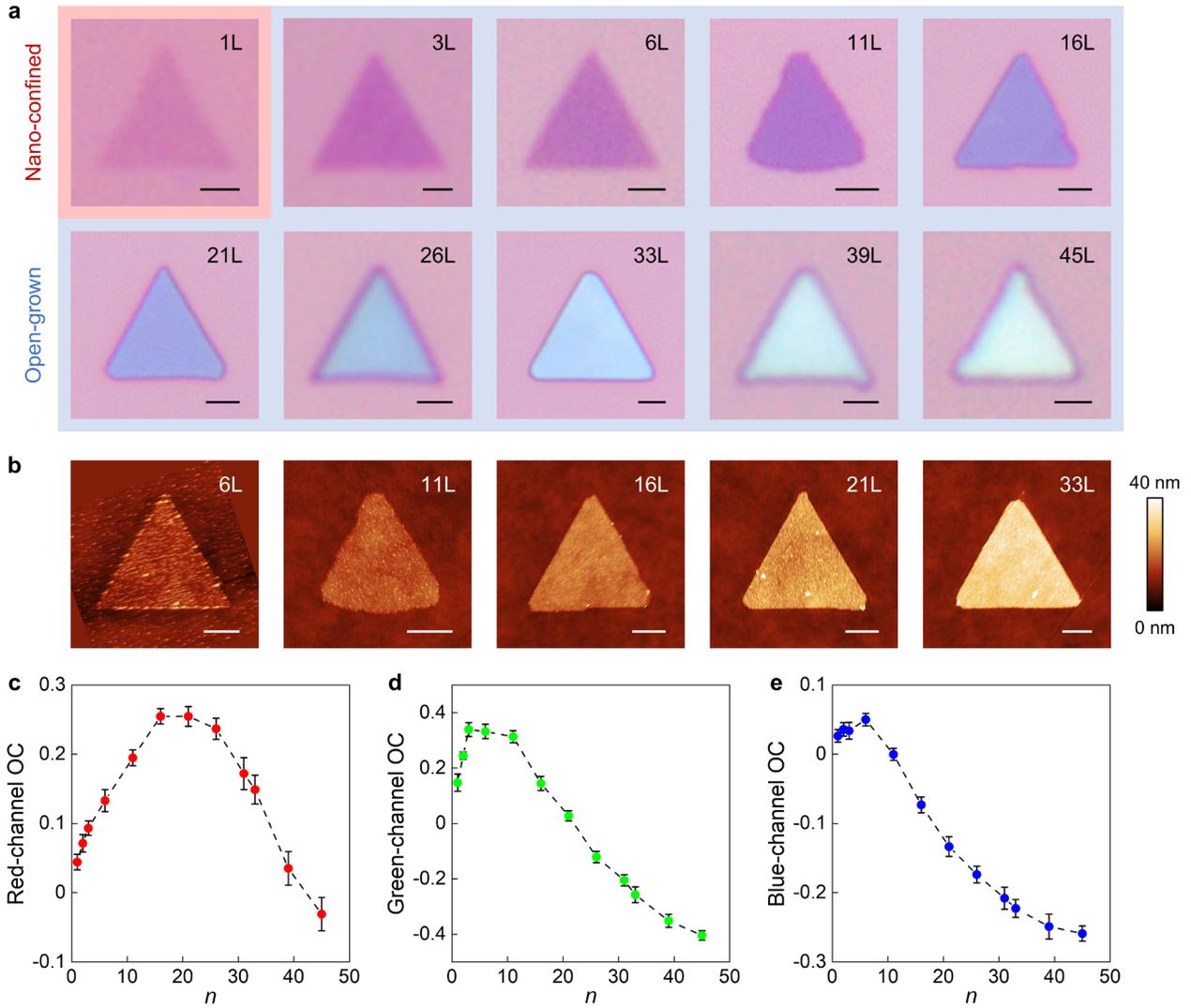

**Fig. S5 | Determination of the thickness of NbSe$_2$ crystals using optical contrast. a**, A series of optical micrographs showing NbSe$_2$ crystals with varying thicknesses and colors. **b**, Typical atomic force microscope (AFM) topographic images confirming the variation in thickness. Scale bars in **a** and **b**: 1 μm (1L, 3L, 6L, 26L, 39L, and 45L); 2 μm (11L, 16L, 21L, and 33L). **c-e**, Dependences of optical contrast (OC) on the crystal thickness (i.e., the number of layers, $n$) for the red (**c**), green (**d**), and blue (**e**) channels. OC was calculated as $(c_1 - c_2)/c_1$, where $c_1$ and $c_2$ are the color values of the SiO$_2$ substrate and the NbSe$_2$ crystal, respectively. The OC-thickness dependences of the 3 channels were used jointly to determine the thicknesses of NbSe$_2$ crystals[17-19], which enabled rapid statistical characterization of ~820 crystals (Fig. 1g) without requiring AFM for direct thickness measurement.

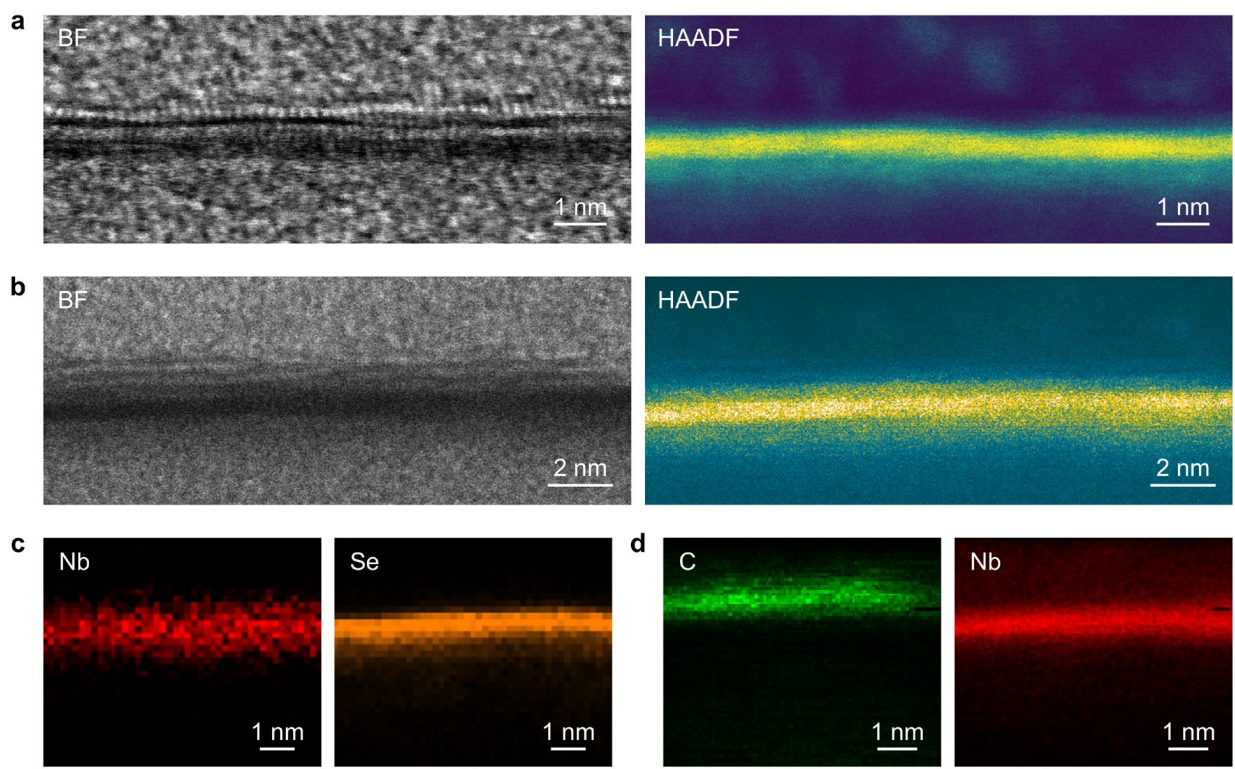

**Fig. S6 | Cross-sectional STEM characterizations of graphene-confined NbSe$_2$. a,b**, Additional STEM images which further evidence the nano-confined growth of NbSe$_2$ underneath graphene. **a**, Additional Sample 1. Similar to Fig. 1c,d in the main text, the BF image shows the stacking of two monolayers with dark contrast and different thicknesses. The thinner and thicker layers on the upper and lower sides are identified to be graphene and NbSe$_2$, respectively. The corresponding HAADF image further verifies this identification. The NbSe$_2$ monolayer is revealed as the bright stripe while the graphene monolayer can be hardly visible in the HAADF image, which is consistent with their large difference in atomic number. **b**, Additional Sample 2. Similarly, the dark stripe in the BF image together with the bright stripe in the HAADF image led to the recognition of NbSe$_2$ monolayer. The lattice fringes above the NbSe$_2$ layer, more clearly observed in the BF image, originate from a few-layer graphene flake. **c,d** EELS characterizations of Additional Sample 2. **c**, Nb and Se mappings verifying the desired composition of NbSe$_2$. **d**, C and Nb mappings confirming the nano-confined growth of NbSe$_2$ (red stripe in the Nb mapping) underneath graphene (green stripe in the C mapping).

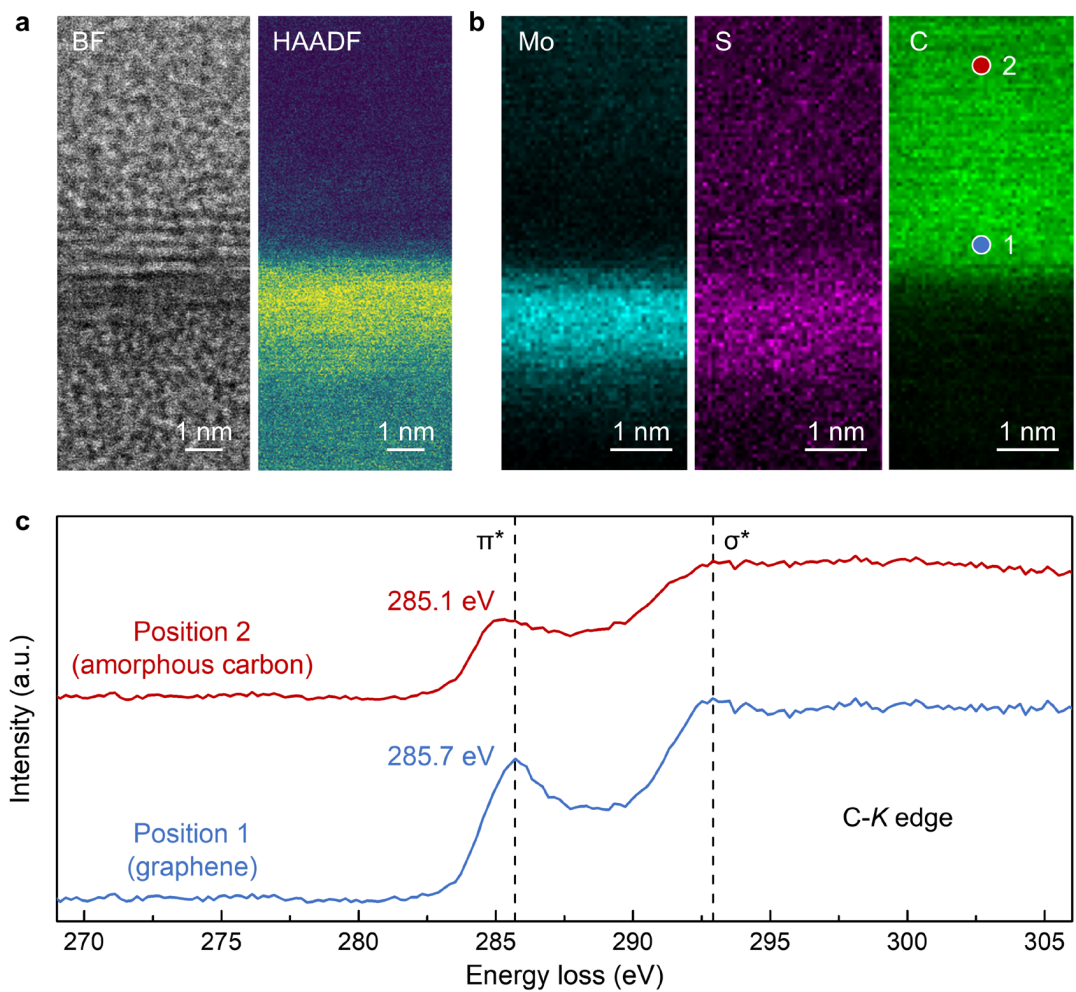

**Fig. S7 | Cross-sectional STEM characterizations of graphene-confined MoS$_2$. a**, STEM images revealing the nano-confined growth of MoS$_2$ underneath graphene. The MoS$_2$ monolayer appears as a bright stripe in the HAADF image, while the BF image shows lattice fringes from a few-layer graphene flake and an amorphous-carbon protective layer on top of the graphene flake. **b**, EELS elemental mappings further confirming the nano-confined growth of MoS$_2$ underneath graphene. The bright stripes in the Mo and S mappings indicate the MoS$_2$ monolayer. The C mapping shows signals from both the graphene flake and the amorphous-carbon protective layer. **c**, C-*K* edge EELS spectra from graphene and amorphous carbon (Positions 1 and·2 in **b**, respectively). The dashed lines indicate the π* and σ* peaks. Graphene exhibits a shaper π* peak at a higher energy loss (285.7 eV) compared to amorphous carbon (285.1 eV)[20,21], thereby enabling clear separation of these two types of carbon using the π* peak.

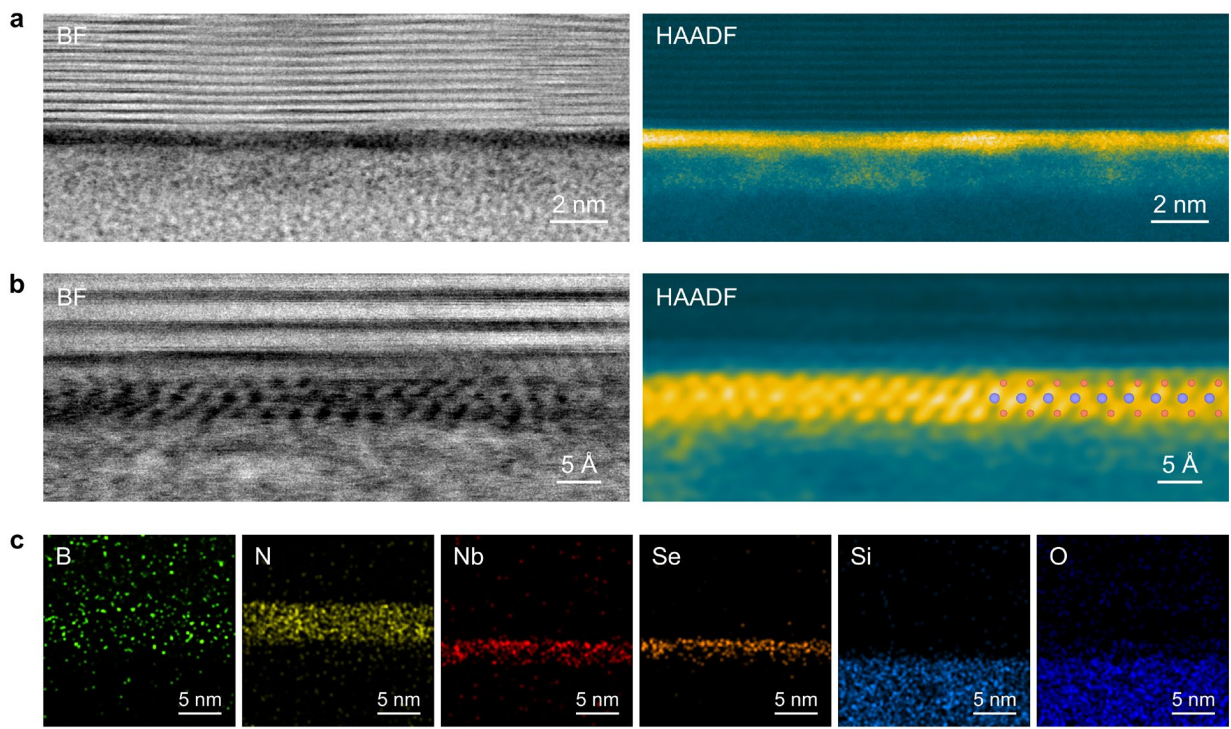

**Fig. S8 | Cross-sectional STEM characterizations of hBN-confined NbSe$_2$. a**, STEM images revealing the nano-confined growth of NbSe$_2$ underneath hBN. The NbSe$_2$ monolayer appears as a bright stripe in the HAADF image, while the lattice fringes above correspond to a multilayer hBN flake. **b**, Close-up STEM images clearly resolving the atomic arrangement of the NbSe$_2$ monolayer. The atomic model along the <100> zone axis is overlaid on the HAADF image, with the purple and orange balls representing the Nb and Se atoms, respectively. **c**, Energy-dispersive X-ray spectroscopy (EDS) elemental mappings further confirming the nano-confined growth of NbSe$_2$ underneath hBN.

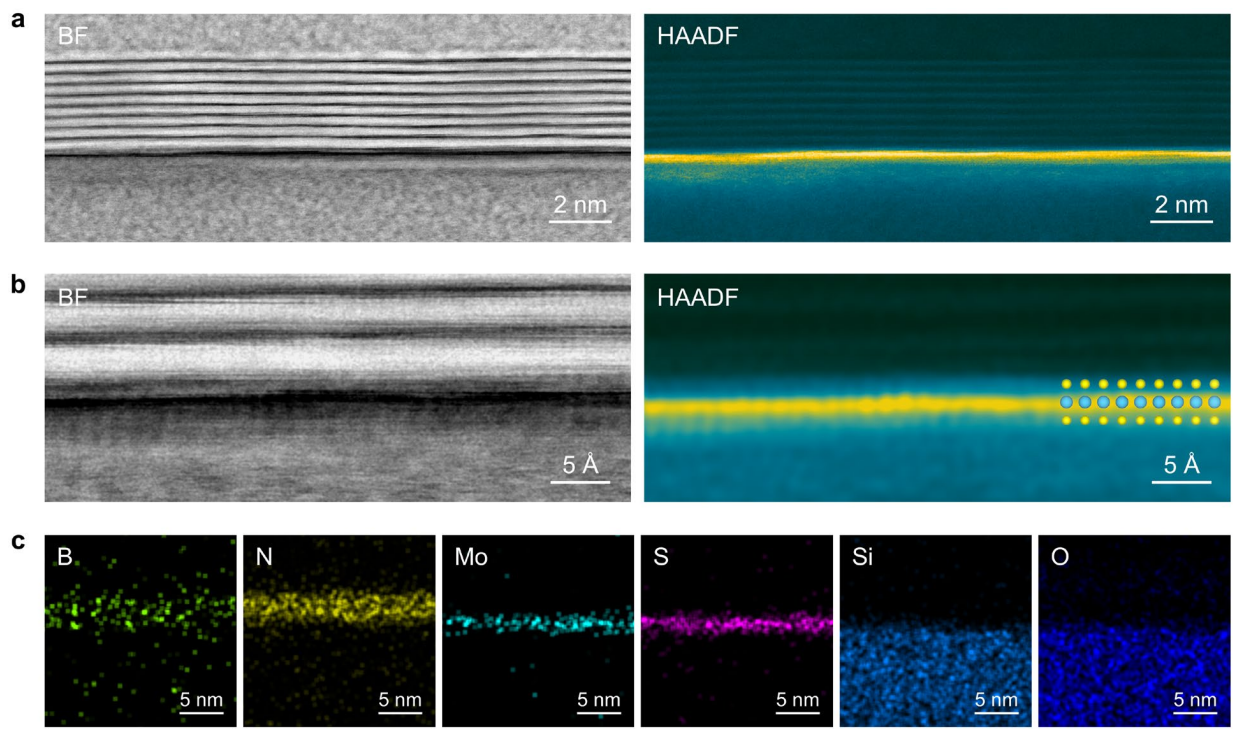

**Fig. S9 | Cross-sectional STEM characterizations of hBN-confined MoS₂. a**, STEM images revealing the nano-confined growth of MoS$_2$ underneath hBN. The MoS$_2$ monolayer appears as a bright stripe in the HAADF image, while the lattice fringes above correspond to a few-layer hBN flake. **b**, Close-up STEM images clearly resolving the atomic arrangement of the MoS$_2$ monolayer. The atomic model along the <210> zone axis is overlaid on the HAADF image, with the cyan and yellow balls representing the Mo and S atoms, respectively. **c**, Energy-dispersive X-ray spectroscopy (EDS) elemental mappings further confirming the nano-confined growth of MoS$_2$ underneath hBN.

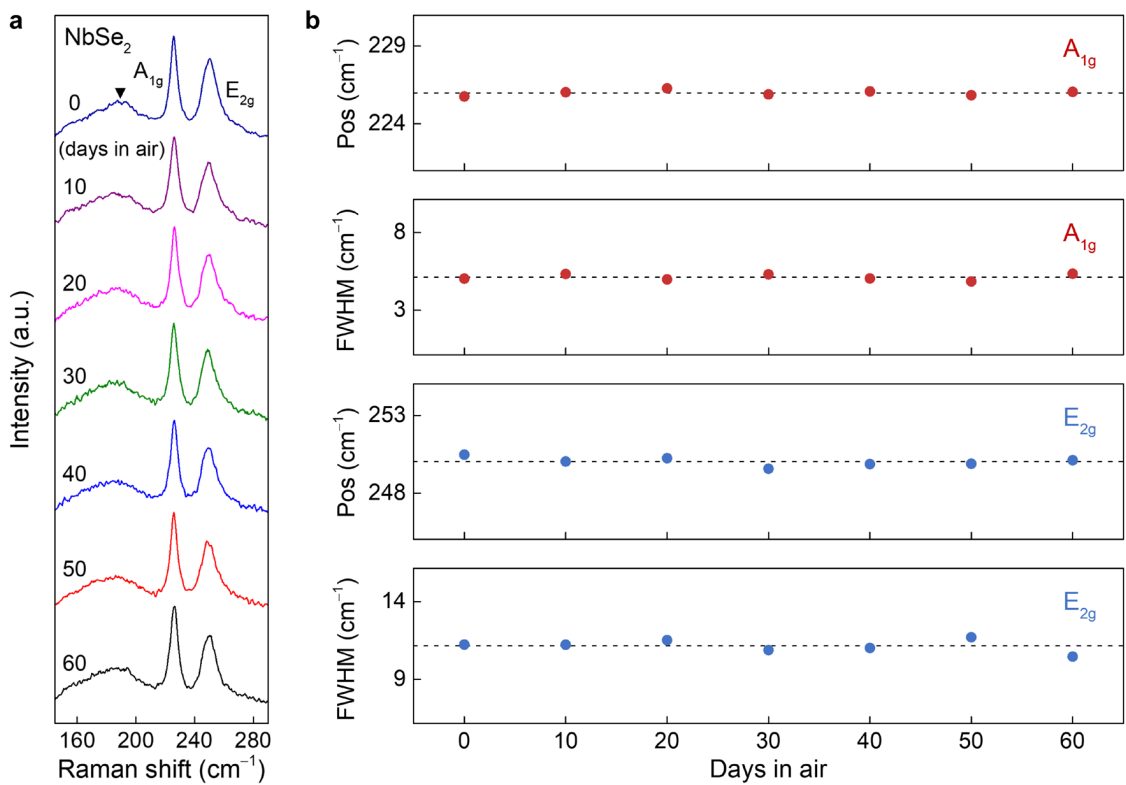

**Fig. S10 | Air stability of graphene-confined NbSe₂. a**, Raman spectra of the graphene-confined NbSe₂ monolayers, acquired over 60 days of air exposure after synthesis. The typical $A_{1g}$ and $E_{2g}$ bands of NbSe₂ along with the soft mode (indicated by ▼) are clearly visible in all spectra, demonstrating well preserved crystal quality of the NbSe₂ monolayers enabled by in situ graphene encapsulation. **b**, Peak positions (Pos) and full widths at half maximum (FWHMs) of the $A_{1g}$ and $E_{2g}$ bands extracted from **a**. The dashed lines indicate the average values. Both parameters remain essentially unchanged over the entire 60-day period, further demonstrating exceptional air stability of the NbSe₂ monolayers.

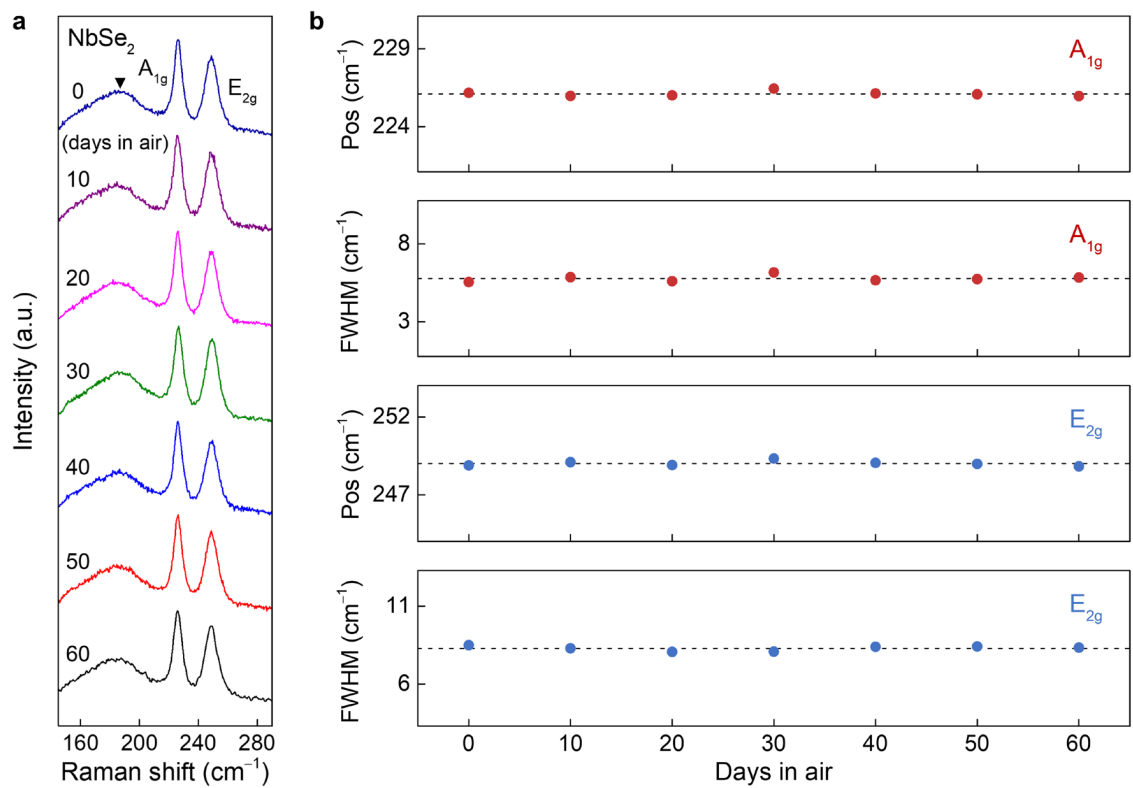

**Fig. S11 | Air stability of hBN-confined NbSe$_2$. a**, Raman spectra of the hBN-confined NbSe$_2$ monolayers, acquired over 60 days of air exposure after synthesis. The typical A$_{1g}$ and E$_{2g}$ bands of NbSe$_2$ along with the soft mode (indicated by ▼) are clearly visible in all spectra, demonstrating well preserved crystal quality of the NbSe$_2$ monolayers enabled by in situ hBN encapsulation. **b**, Peak positions (Pos) and full widths at half maximum (FWHMs) of the A$_{1g}$ and E$_{2g}$ bands extracted from **a**. The dashed lines indicate the average values. Both parameters remain essentially unchanged over the entire 60-day period, further demonstrating exceptional air stability of the NbSe$_2$ monolayers.

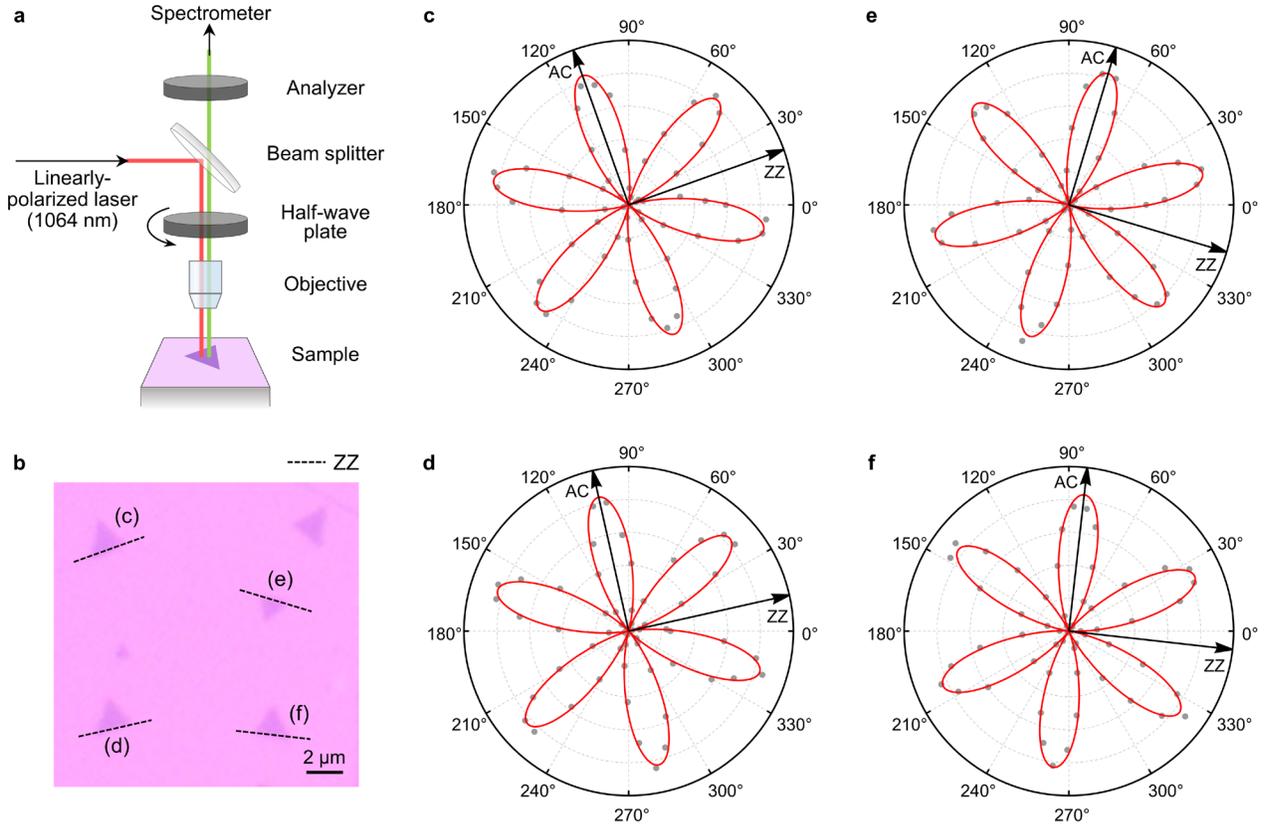

**Fig. S12 | Lattice orientation of nano-confined MoS₂ determined using SHG. a**, Schematic diagram illustrating the setup for the polarization-resolved SHG measurement. A linearly-polarized fundamental beam at 1064 nm was focused on the sample at normal incidence, and the SHG signal was collected in reflection geometry. The polarizations of the fundamental beam as well as the SHG signal were simultaneously rotated by a half-wave plate mounted on a motorized rotation stage. The SHG signal was then passed through an analyzer (i.e., a linear polarizer) before being coupled into a charge-coupled device (CCD) spectrometer. The SHG intensity of a MoS₂ monolayer follows[22,23]

$$I_{2\omega} \propto \sin^2[3(\theta - \theta_{zz}) - \Phi], \tag{S1}$$

where $\theta$ is the polarization angle of the fundamental beam, $\theta_{zz}$ denotes the zigzag (ZZ) direction of MoS₂ lattice, and $\Phi$ is a phase depending on the detailed configurations of the setup and was tuned to be zero. **b**, Optical micrograph of the nano-confined MoS₂ monolayers under measurements. **c-f**, $I_{2\omega}(\theta)$ of four different MoS₂ monolayers. All polarization dependences exhibit the 6-fold variations which can be well fitted by equation (S1) (red curves). The lattice orientations determined from the fittings are indicated by the arrows in **c-f**. Further correlating the lattice orientations with the optical micrograph (**b**) clearly indicates the ZZ edge structure of the MoS₂ monolayers (dashed lines in **b**)[24].

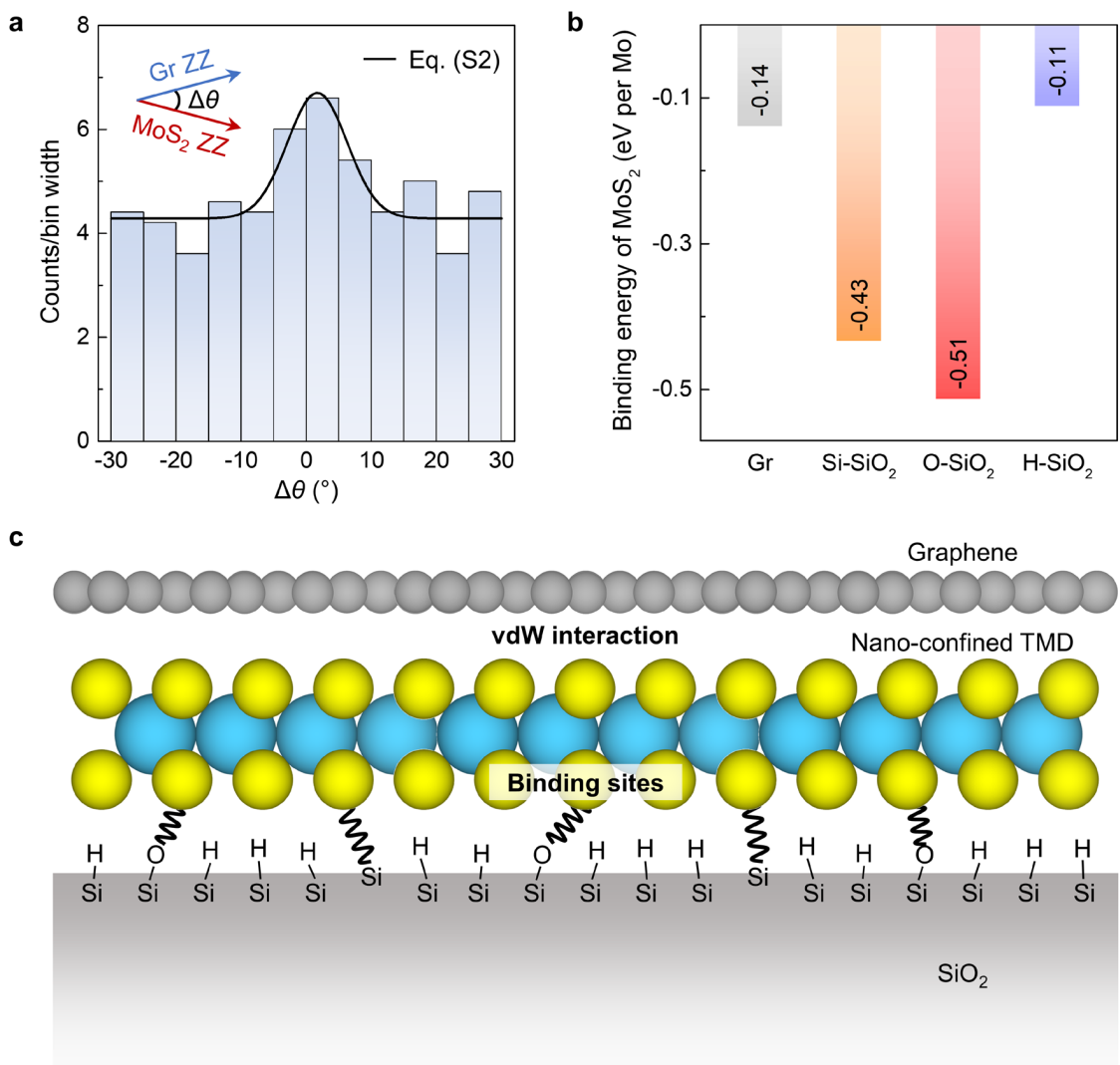

**Fig. S13 | Relative lattice orientation between nano-confined MoS$_2$ and graphene. a**, Histogram showing the relative lattice orientation between the nano-confined MoS$_2$ monolayer and graphene. The lattice orientations of MoS$_2$ and graphene were individually determined using the polarization-resolved SHG spectroscopy (Fig. S12) and the scanning tunneling microscopy (STM)[25], respectively. The relative lattice orientation was defined as the angle $\Delta\theta$ between the ZZ directions of MoS$_2$ and graphene (inset of **a**). Evidently, $\Delta\theta$ exhibits a small peak around 0°, indicating that the MoS$_2$ lattice tends to align with the graphene lattice, an epitaxial behavior that has been commonly reported for the open growth of TMD monolayers on graphene[26-29]. Besides the peak, there also exists a constant background which demonstrates the existence of un-aligned MoS$_2$ lattices with random orientations. The following model was proposed to describe the observed peak-plus-background behaviors of $\Delta\theta$

$$n = N \times \left[\delta \times \text{PDF}(\mu,\sigma) + (1-\delta) \times \frac{1}{60}\right], \tag{S2}$$

where $n$ is the histogram counts normalized by the bin width, $N$ is the total number of samples, $\delta$ is the proportion of the aligned MoS$_2$ lattices, PDF is the probability density function of the Gaussian distribution with peak position $\mu$ and standard deviation $\sigma$. Fitting the data with equation (S2) yields $\mu = 1.7°$, $\sigma = 4.6°$, and $\delta = 0.10$ (black curve in **a**). Accordingly, only 10% of the MoS$_2$ lattices align with that of the graphene, while the remaining 90% are randomly oriented. Such a partially-aligned texture can be well explained by invoking the "competing" interactions at the two sides of the nano-confined MoS$_2$ monolayer. On one hand, the MoS$_2$/graphene vdW interaction serves as the driving force for lattice alignment[28]; on the other hand, the interaction between MoS$_2$ and amorphous SiO$_2$ considerably disturbs the alignment. **b**, Binding energy of a MoS$_2$ monolayer when being interfaced with graphene and SiO$_2$, calculated using DFT. It is clear that the MoS$_2$ monolayer exhibits stronger binding strengths with the unsaturated SiO$_2$ surfaces terminated by Si and O, while a much weaker binding strength with the inner graphene surface. Therefore, it is likely that interfacial binding sites exist between MoS$_2$ and SiO$_2$, which greatly disturbs the lattice alignment and leads to the partially-aligned texture of the MoS$_2$ lattices. **c**, Schematic diagram illustrating the "double-sided" scheme of interaction for the nano-confined TMD monolayer. The springs indicate the interfacial binding sites.

**Table S2 | Atomic models of different growth schemes (I).** Atomic models of the nano-confined growth with the graphene capping layer (graphene/TMD/SiO$_2$), the open growth on SiO$_2$ (TMD/SiO$_2$), and the open growth on graphene (TMD/graphene/SiO$_2$), with various SiO$_2$ terminations to model the amorphous substrate[30]. Atom colors: C, gray; Si, blue; O, red; H, white; Nb, purple; Se, orange; Mo, cyan; S, yellow.

| TMD | SiO$_2$ termination | Growth scheme | | |
| --- | --- | --- | --- | --- |
| | | Open growth on SiO$_2$ | Open growth on Gr | Nano-confined growth |
| NbSe$_2$ | O dangling bond | 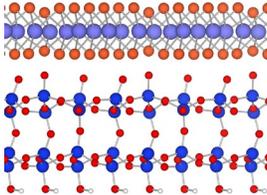 | 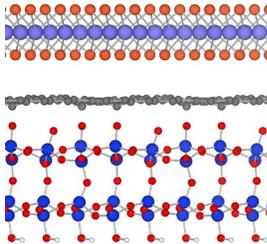 | 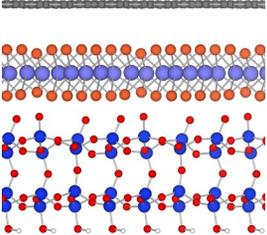 |
| NbSe$_2$ | Si dangling bond | 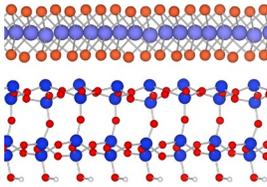 | 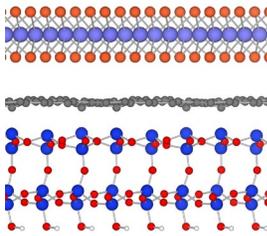 | 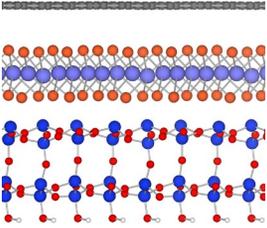 |
| NbSe$_2$ | Si dangling bond partially passivated by H | 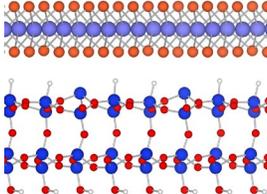 | 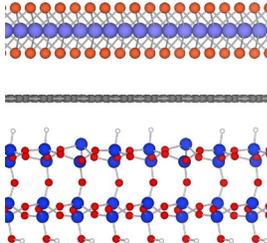 | 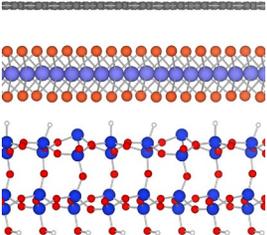 |

| TMD | SiO$_2$ termination | Growth scheme | | |
| --- | --- | --- | --- | --- |
| | | Open growth on SiO$_2$ | Open growth on Gr | Nano-confined growth |
| MoS$_2$ | O dangling bond | 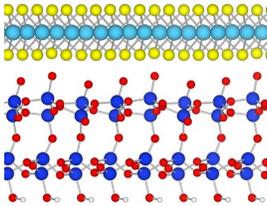 | 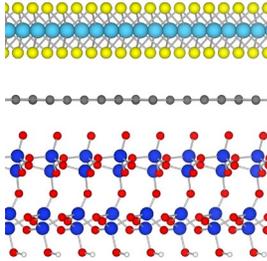 | 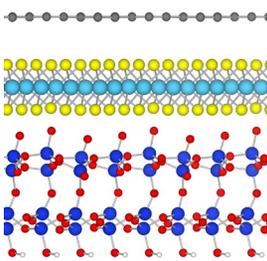 |
| MoS$_2$ | Si dangling bond | 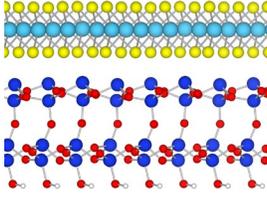 | 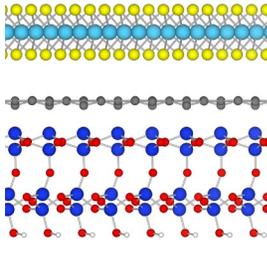 | 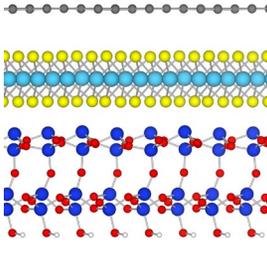 |
| MoS$_2$ | Si dangling bond partially passivated by H | 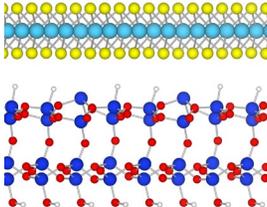 | 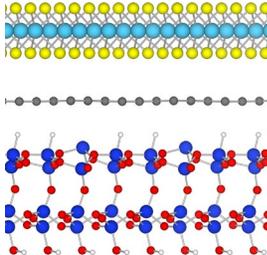 | 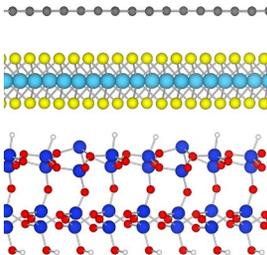 |

**Table S3 | Binding energy of TMD monolayers for different growth schemes (I).** The binding energy was calculated as $E_{binding} = E_{total} - E_{TMD} - \sum E_{neighbor}$ (ref. [31]), where $E_{total}$ is the total energy of the adsorbing system shown in Table S2, $E_{TMD}$ is the energy of the isolated TMD monolayer, and $E_{neighbor}$ is the energy of the isolated neighboring layer defined in the parentheses below. The nano-confined TMD monolayers exhibit the most negative $E_{binding}$ regardless of $SiO_2$ terminations, demonstrating energetically favored growth under the nano-confinement.

| TMD | SiO$_2$ termination | Growth scheme | Energy (eV per metal atom) | | | |
| --- | --- | --- | --- | --- | --- | --- |
| | | | $E_{total}$ | $E_{TMD}$ | $E_{neighbor}$ | $E_{binding}$ |
| NbSe$_2$ | O dangling bond | Open growth on SiO$_2$ | −63.67 | | −42.58 (SiO$_2$) | −0.60 |
| | | Open growth on Gr | −98.09 | −20.49 | −77.34 (Gr/SiO$_2$) | −0.26 |
| | | Nano-confined growth | −98.26 | | −34.41 (Gr), −42.58 (SiO$_2$) | −0.78 |
| NbSe$_2$ | Si dangling bond | Open growth on SiO$_2$ | −60.58 | | −39.29 (SiO$_2$) | −0.80 |
| | | Open growth on Gr | −94.79 | −20.49 | −74.10 (Gr/SiO$_2$) | −0.20 |
| | | Nano-confined growth | −95.16 | | −34.41 (Gr), −39.29 (SiO$_2$) | −0.97 |
| NbSe$_2$ | Si dangling bond partially passivated by H | Open growth on SiO$_2$ | −61.35 | | −40.70 (SiO$_2$) | −0.16 |
| | | Open growth on Gr | −95.88 | −20.49 | −75.20 (Gr/SiO$_2$) | −0.19 |
| | | Nano-confined growth | −95.93 | | −34.41 (Gr), −40.70 (SiO$_2$) | −0.34 |

| TMD | SiO$_2$ termination | Growth scheme | Energy (eV per metal atom) | | | |
|---|---|---|---|---|---|---|
| | | | $E_{total}$ | $E_{TMD}$ | $E_{neighbor}$ | $E_{binding}$ |
| MoS$_2$ | O dangling bond | Open growth on SiO$_2$ | −65.38 | | −42.61 (SiO$_2$) | −0.51 |
| | | Open growth on Gr | −97.13 | −22.26 | −74.74 (Gr/SiO$_2$) | −0.14 |
| | | Nano-confined growth | −97.25 | | −31.71 (Gr), −42.61 (SiO$_2$) | −0.68 |
| MoS$_2$ | Si dangling bond | Open growth on SiO$_2$ | −62.01 | | −39.32 (SiO$_2$) | −0.43 |
| | | Open growth on Gr | −93.72 | −22.26 | −71.29 (Gr/SiO$_2$) | −0.17 |
| | | Nano-confined growth | −93.86 | | −31.71 (Gr), −39.32 (SiO$_2$) | −0.58 |
| MoS$_2$ | Si dangling bond partially passivated by H | Open growth on SiO$_2$ | −63.03 | | −40.67 (SiO$_2$) | −0.11 |
| | | Open growth on Gr | −94.90 | −22.26 | −72.48 (Gr/SiO$_2$) | −0.16 |
| | | Nano-confined growth | −94.89 | | −31.71 (Gr), −40.67 (SiO$_2$) | −0.26 |

**Table S4 | Atomic models of different growth schemes (II).** Atomic models of the nano-confined growth with the hBN capping layer (hBN/TMD/SiO$_2$), the open growth on SiO$_2$ (TMD/SiO$_2$), and the open growth on hBN (TMD/hBN/SiO$_2$), with various SiO$_2$ terminations to model the amorphous substrate[30]. Atom colors: B, green; N, dark yellow; Si, blue; O, red; H, white; Nb, purple; Se, orange; Mo, cyan; S, yellow.

| TMD | SiO$_2$ termination | Growth scheme | | |
| --- | --- | --- | --- | --- |
| | | Open growth on SiO$_2$ | Open growth on hBN | Nano-confined growth |
| NbSe$_2$ | O dangling bond | 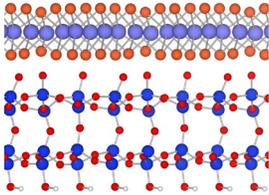 | 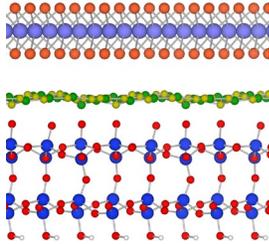 | 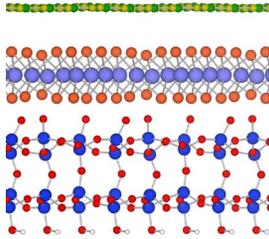 |
| NbSe$_2$ | Si dangling bond | 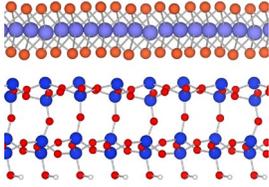 | 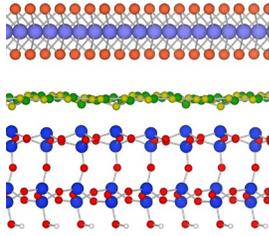 | 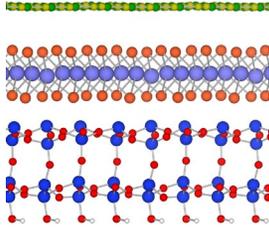 |
| NbSe$_2$ | Si dangling bond partially passivated by H | 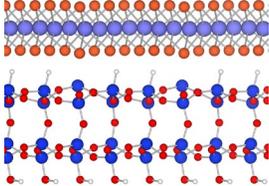 | 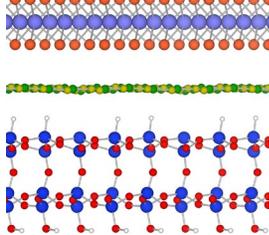 | 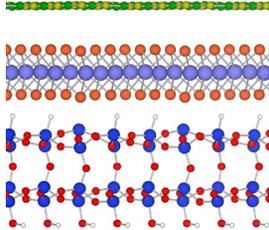 |

| TMD | SiO$_2$ termination | Growth scheme | | |
| --- | --- | --- | --- | --- |
| | | Open growth on SiO$_2$ | Open growth on hBN | Nano-confined growth |
| MoS$_2$ | O dangling bond | 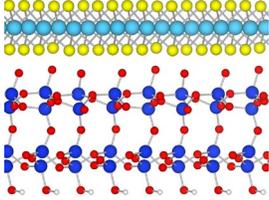 | 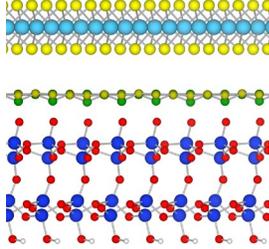 | 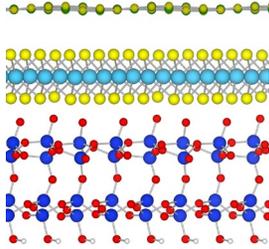 |
| MoS$_2$ | Si dangling bond | 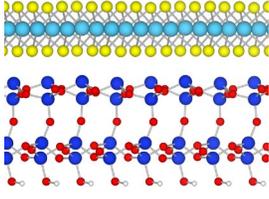 | 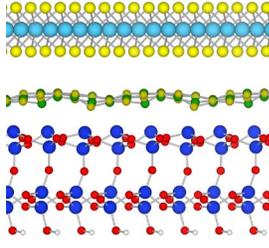 | 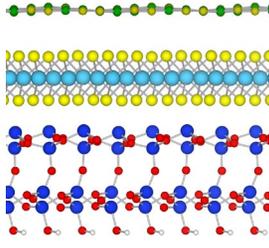 |
| MoS$_2$ | Si dangling bond partially passivated by H | 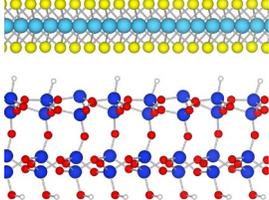 | 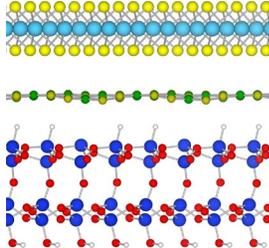 | 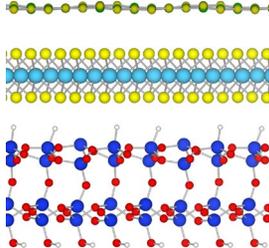 |

**Table S5 | Binding energy of TMD monolayers for different growth schemes (II).** The binding energy was calculated as $E_{binding} = E_{total} - E_{TMD} - \sum E_{neighbor}$ (ref. [31]), where $E_{total}$ is the total energy of the adsorbing system shown in Table S4, $E_{TMD}$ is the energy of the isolated TMD monolayer, and $E_{neighbor}$ is the energy of the isolated neighboring layer defined in the parentheses below. The nano-confined TMD monolayers exhibit the most negative $E_{binding}$ regardless of $SiO_2$ terminations, demonstrating energetically favored growth under the nano-confinement.

| TMD | SiO$_2$ termination | Growth scheme | Energy (eV per metal atom) | | | |
|---|---|---|---|---|---|---|
| | | | $E_{total}$ | $E_{TMD}$ | $E_{neighbor}$ | $E_{binding}$ |
| NbSe$_2$ | O dangling bond | Open growth on SiO$_2$ | −65.01 | | −43.86 (SiO$_2$) | −0.85 |
| | | Open growth on hBN | −97.77 | −20.30 | −77.25 (hBN/SiO$_2$) | −0.21 |
| | | Nano-confined growth | −98.09 | | −32.87 (hBN), −43.86 (SiO$_2$) | −1.06 |
| NbSe$_2$ | Si dangling bond | Open growth on SiO$_2$ | −62.06 | | −40.89 (SiO$_2$) | −0.86 |
| | | Open growth on hBN | −94.72 | −20.30 | −74.18 (hBN/SiO$_2$) | −0.23 |
| | | Nano-confined growth | −95.12 | | −32.87 (hBN), −40.89 (SiO$_2$) | −1.05 |
| NbSe$_2$ | Si dangling bond partially passivated by H | Open growth on SiO$_2$ | −63.16 | | −42.57 (SiO$_2$) | −0.29 |
| | | Open growth on hBN | −96.02 | −20.30 | −75.51 (hBN/SiO$_2$) | −0.21 |
| | | Nano-confined growth | −96.22 | | −32.87 (hBN), −42.57 (SiO$_2$) | −0.47 |

| TMD | SiO2 termination | Growth scheme | Energy (eV per metal atom) | | | |
| --- | --- | --- | --- | --- | --- | --- |
| | | | $E_{total}$ | $E_{TMD}$ | $E_{neighbor}$ | $E_{binding}$ |
| MoS2 | O dangling bond | Open growth on SiO2 | −67.00 | | −43.94 (SiO2) | −0.68 |
| | | Open growth on hBN | −97.36 | −22.38 | −74.80 (hBN/SiO2) | −0.18 |
| | | Nano-confined growth | −97.37 | | −30.20 (hBN), −43.94 (SiO2) | −0.85 |
| MoS2 | Si dangling bond | Open growth on SiO2 | −63.86 | | −41.03 (SiO2) | −0.44 |
| | | Open growth on hBN | −94.16 | −22.38 | −71.60 (hBN/SiO2) | −0.18 |
| | | Nano-confined growth | −94.22 | | −30.20 (hBN), −41.03 (SiO2) | −0.60 |
| MoS2 | Si dangling bond partially passivated by H | Open growth on SiO2 | −65.13 | | −42.67 (SiO2) | −0.08 |
| | | Open growth on hBN | −95.53 | −22.38 | −72.97 (hBN/SiO2) | −0.17 |
| | | Nano-confined growth | −95.50 | | −30.20 (hBN), −42.67 (SiO2) | −0.24 |

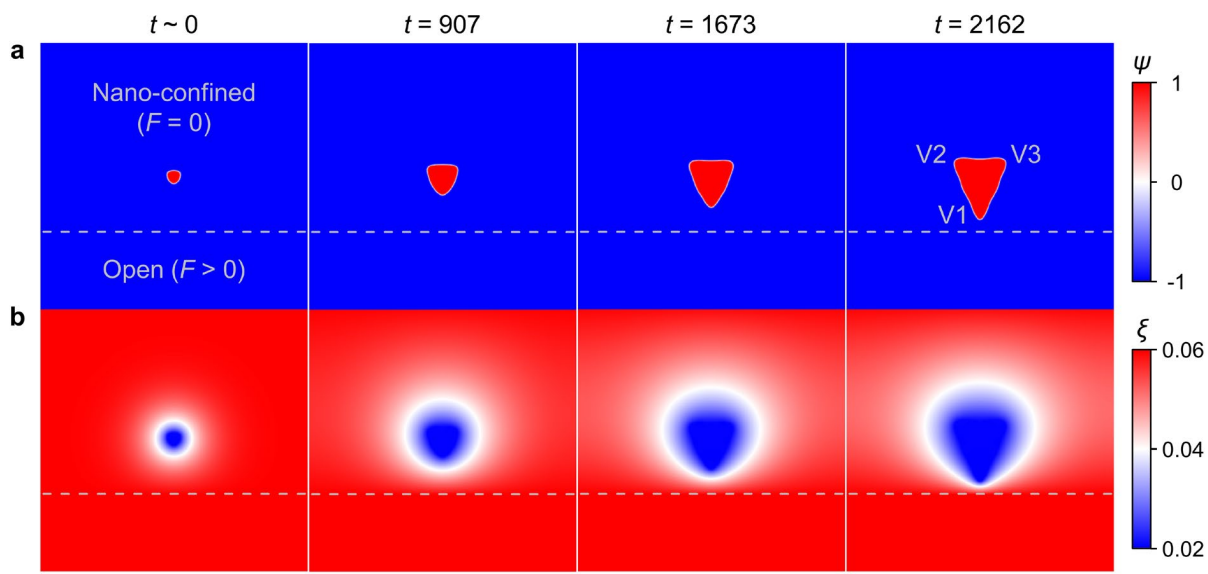

**Fig. S14 | Phase-field simulations incorporating edge intercalation.** The simulation box consists of two regions, the upper nano-confined interface with deposition flux $F = 0$ and the lower open $SiO_2$ substrate with $F > 0$, separated by a dashed line representing the edge of the capping layer. The edge intercalation implies that precursors deposited in the open region can diffuse across the dashed line into the confined region. **a**, Evolution of growth shape simulated by an order parameter $\psi$ ranging from $-1$ (ungrown) to 1 (fully grown)[32,33]. A small circular nucleus with $\psi = 1$ is preset in the confined region and evolves into a triangular domain at $t = 907$. Subsequently, a significant overgrowth develops at the edge-facing vertex V1, increasing the degree of growth asymmetry (DGA) from 0.05 at $t = 907$ to 0.22 at $t = 2162$, with the overgrowth direction toward the edge. These results clearly reproduce the observed asymmetric growth behavior (Figs. 2b and S2), confirming the edge-intercalation mechanism. **b**, Evolution of precursor concentration field ($\xi$). $\xi$ is notably higher near the edge-facing vertex V1 than near V2 and V3, indicating that V1 acquires precursors more efficiently than the other vertices. Consequently, overgrowth preferentially occurs at V1, leading to asymmetric growth toward the edge.

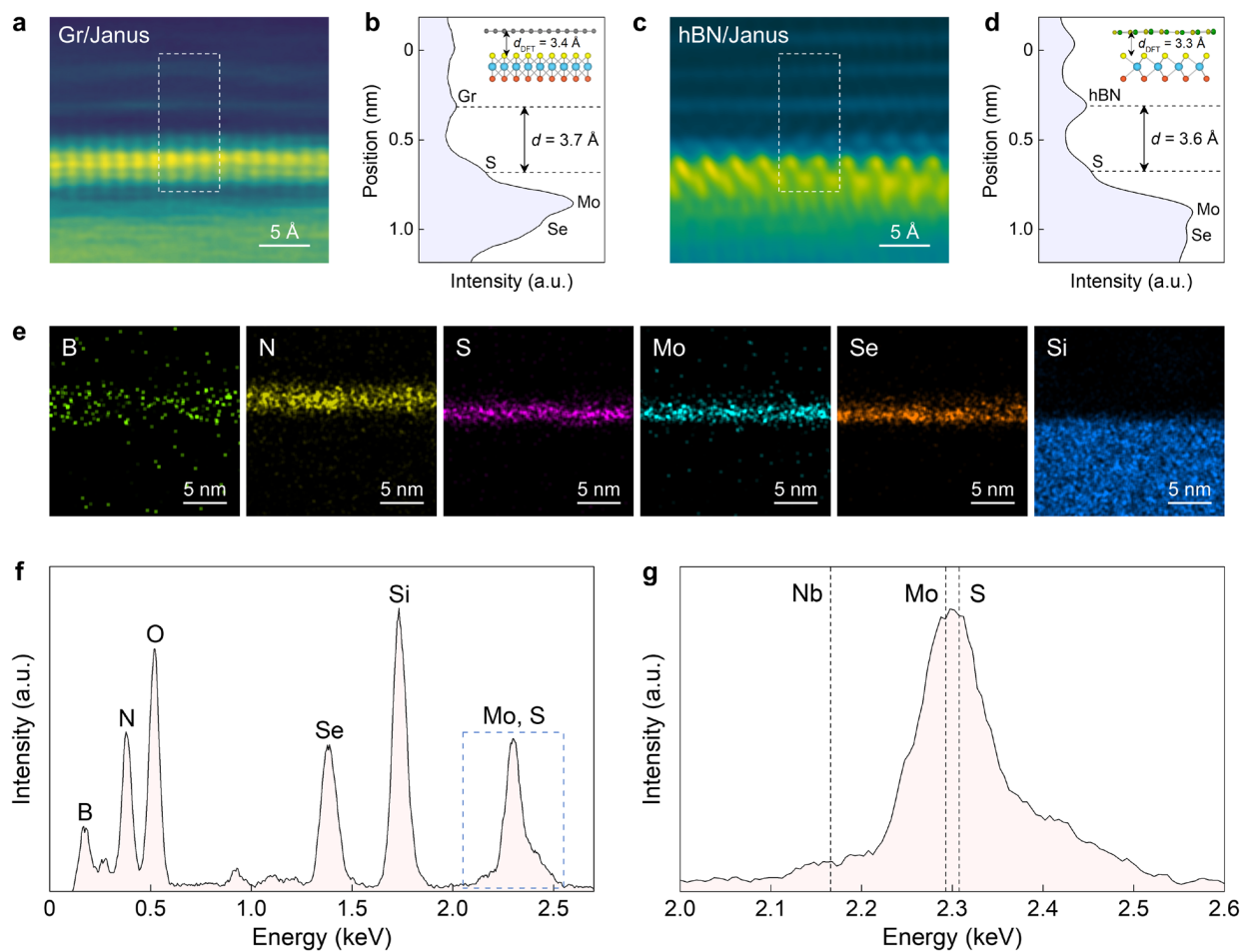

**Fig. S15 | Cross-sectional STEM characterizations of the Janus MoSSe monolayers. a,c**, Atomically-resolved HAADF images of the Janus MoSSe monolayers synthesized using graphene (**a**) or hBN (**c**) as the capping layer. **b,d**, Intensity profiles averaged over the dashed regions in **a,c**, respectively. The intensity difference between the lower Se and upper S atoms clearly evidences the polar chalcogen arrangement within the Janus MoSSe monolayers. The vdW gap between graphene (hBN) and the Janus MoSSe monolayer was determined to be 3.7 Å (3.6 Å), in agreement with the DFT-calculated value shown in the inset of **b** (**d**), demonstrating ultraclean vdW interfaces benefiting from the simultaneous integration. The gray, green, dark yellow, cyan, yellow, and orange balls in the insets of **b,d** represent the C, B, N, Mo, S, and Se atoms, respectively. **e**, EDS elemental mappings of the hBN-confined Janus MoSSe monolayer, further corroborating the formation of Janus MoSSe underneath hBN. **f.** Corresponding EDS spectrum. The characteristic peaks of all expected elements are clearly revealed. **g**, Close-up of the dashed region in **f**, showing strong, overlapping peaks of Mo and S and no discernible Nb signals, confirming negligible Nb residue despite the use of NbSe$_2$ precursors.

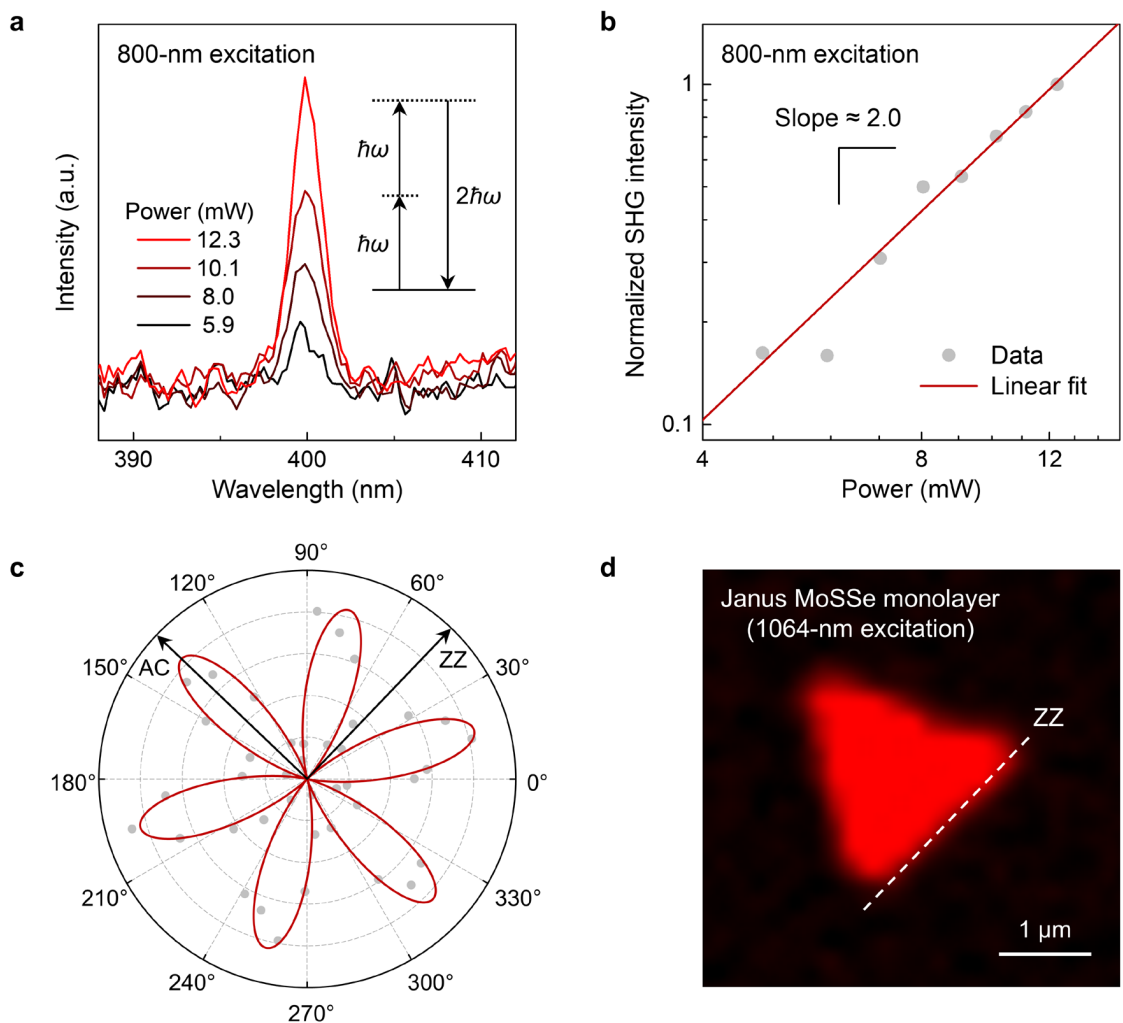

**Fig. S16 | SHG of the Janus MoSSe monolayers. a**, Emergence of SHG at 400 nm when the Janus MoSSe monolayer was excited by the fundamental beam at 800 nm. Inset: energy-level diagram of SHG[34]. The solid and dotted lines denote the ground and virtual levels, respectively. **b**, Laser power dependence of the SHG intensity plotted on the log-log scale. The linear relationship with the slope of 2.0 is expected for SHG[34]. **c**, Polarization-resolved SHG intensity as the function of polarization angle (see Fig. S12a for the setup). At normal incidence and under paraxial approximation, the 6-fold polarization dependence described by equation (S1) is appropriate for both $MoS_2$ and Janus MoSSe monolayers due to their identical in-plane symmetry[35]. As shown in **c**, the polarization dependence of the Janus MoSSe monolayer indeed exhibits the 6-fold variation which is well fitted according to equation (S1) (red curve). The arrows in **c** denote the lattice orientation determined from the fitting. **d**, SHG mapping of the Janus MoSSe monolayer. The uniform SHG intensity confirms the superior

substitution uniformity within the nano-confinement. The ZZ direction extracted from **c** is overlaid on the SHG mapping as the dashed line. Clearly, the Janus MoSSe monolayer exhibits the ZZ edge structure, which agrees with that of the pristine $MoS_2$ monolayers before the substitution (Fig. S12b).

**Note S3 | Discussion regarding the selective intercalation of precursors**

The one-side substitution of the bottom S atoms by the Se atoms indicates the selective intercalation of $NbSe_2$ precursors into the bottom $MoS_2/SiO_2$ interface (inset of Fig. 3g). The underlying reason is two-fold. First, the $SiO_2$ substrate is amorphous and has various surface terminations (Fig. 2a). The alteration of terminating groups and the presence of surface defects could lead to a microscopically rough surface hence abundance of intercalation paths at the $MoS_2/SiO_2$ interface, as opposed to the atomically-smooth vdW interfaces of graphene/$MoS_2$ and hBN/$MoS_2$ that suppress the intercalation kinetically[36]. Furthermore, the precursors have a strong binding strength with the unsaturated $SiO_2$ surface, while weak interactions with the graphene and hBN surfaces owing to their inert nature[37]. Therefore, the intercalation of precursors into the $MoS_2/SiO_2$ interface is also energetically favored.

**Table S6 | Atomic models of different substitution schemes.** Atomic models of one-side and two-side substituted MoS$_{2(1-x)}$Se$_{2x}$ and MoS$_{2x}$Se$_{2(1-x)}$ monolayers at different substitution ratios ($x$), after full relaxation using DFT. Both top and bottom views along the out-of-plane direction, indicated by different shadings, are presented. Cyan, yellow, and orange balls denote Mo, S, and Se, respectively.

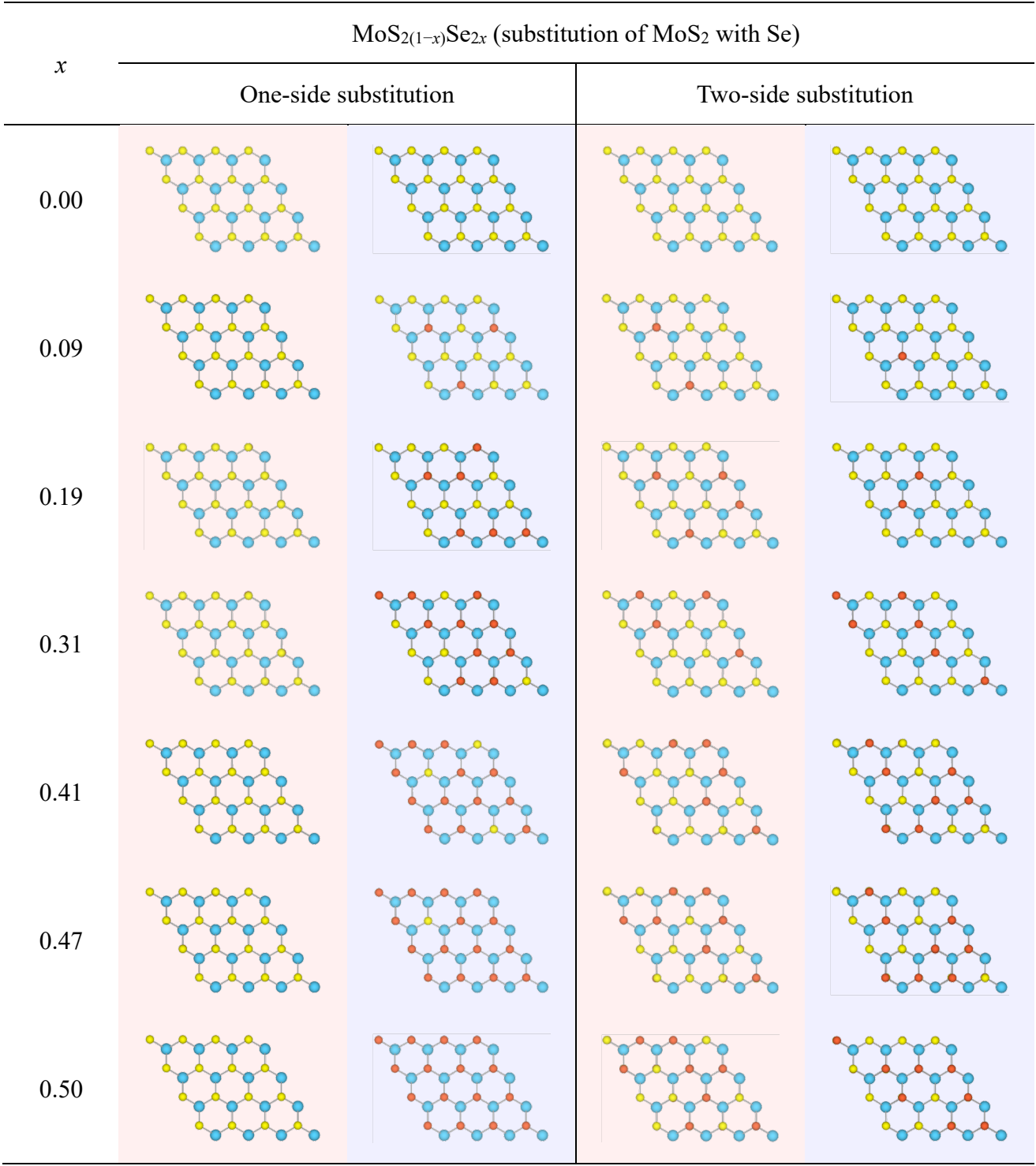

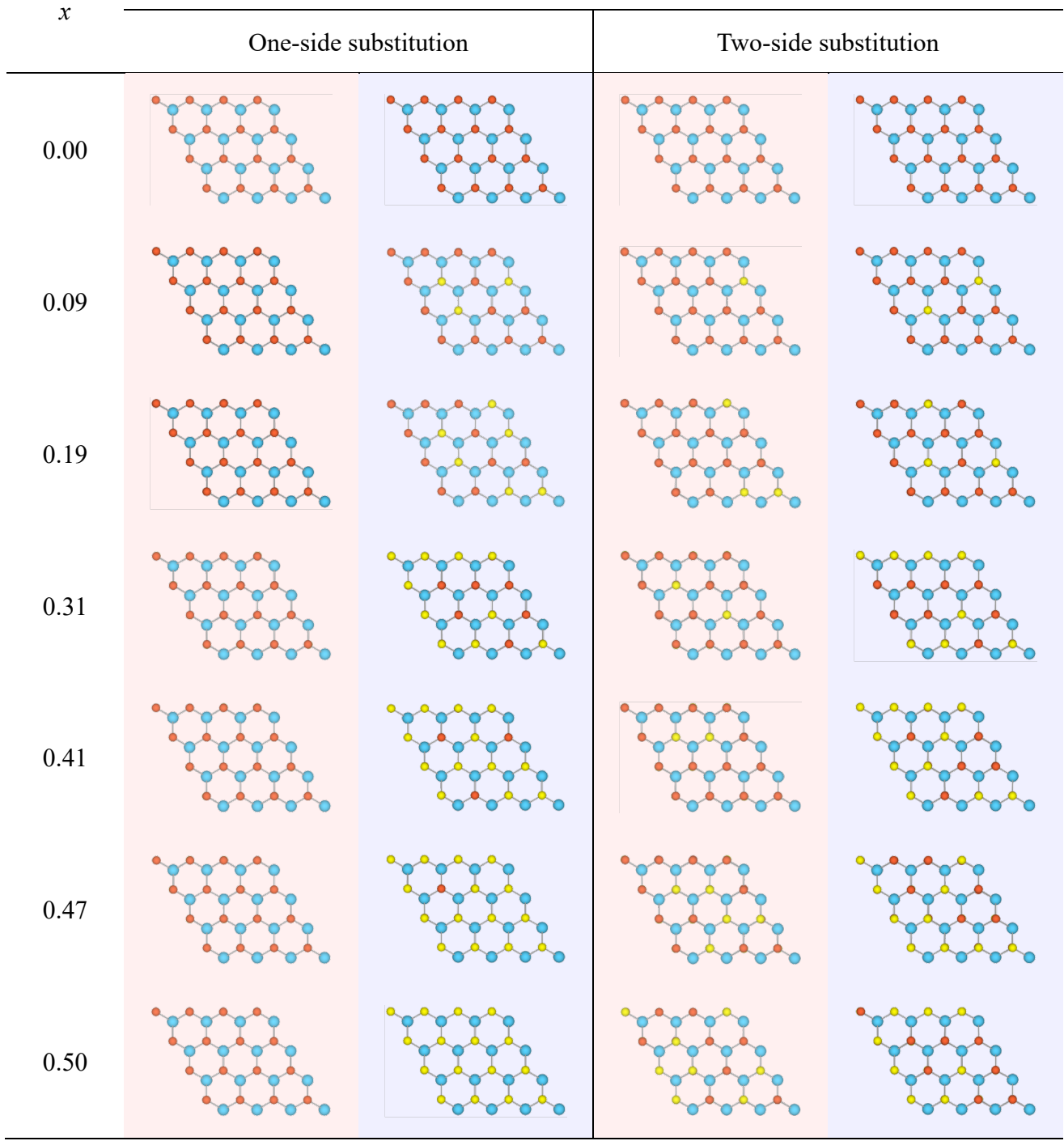

**Table S7 | Energy of the substituted TMD monolayers.** The energy difference between the one-side and two-side substitution products was obtained according to $\Delta E = E_{\text{one-side}} - E_{\text{two-side}}$. Evidently, the one-side substitution is energetically unfavored as the positive $\Delta E$ continuously increases with $x$.

| Substituted TMD | $x$ | Energy (eV per Mo atom) | | |
| --- | --- | --- | --- | --- |
| | | $E_{\text{one-side}}$ | $E_{\text{two-side}}$ | $\Delta E$ |
| MoS$_{2(1-x)}$Se$_{2x}$ | 0.00 | −22.3651 | −22.3651 | 0.0000 |
| | 0.09 | −22.2098 | −22.2106 | 0.0008 |
| | 0.19 | −22.0520 | −22.0559 | 0.0039 |
| | 0.31 | −21.8366 | −21.8503 | 0.0137 |
| | 0.41 | −21.6734 | −21.6936 | 0.0202 |
| | 0.47 | −21.5626 | −21.5893 | 0.0267 |
| | 0.50 | −21.5071 | −21.5385 | 0.0314 |
| MoS$_{2x}$Se$_{2(1-x)}$ | 0.00 | −20.7035 | −20.7035 | 0.0000 |
| | 0.09 | −20.8588 | −20.8592 | 0.0004 |
| | 0.19 | −21.0127 | −21.0165 | 0.0037 |
| | 0.31 | −21.2131 | −21.2202 | 0.0071 |
| | 0.41 | −21.3614 | −21.3711 | 0.0097 |
| | 0.47 | −21.4587 | −21.4830 | 0.0243 |
| | 0.50 | −21.5071 | −21.5385 | 0.0314 |

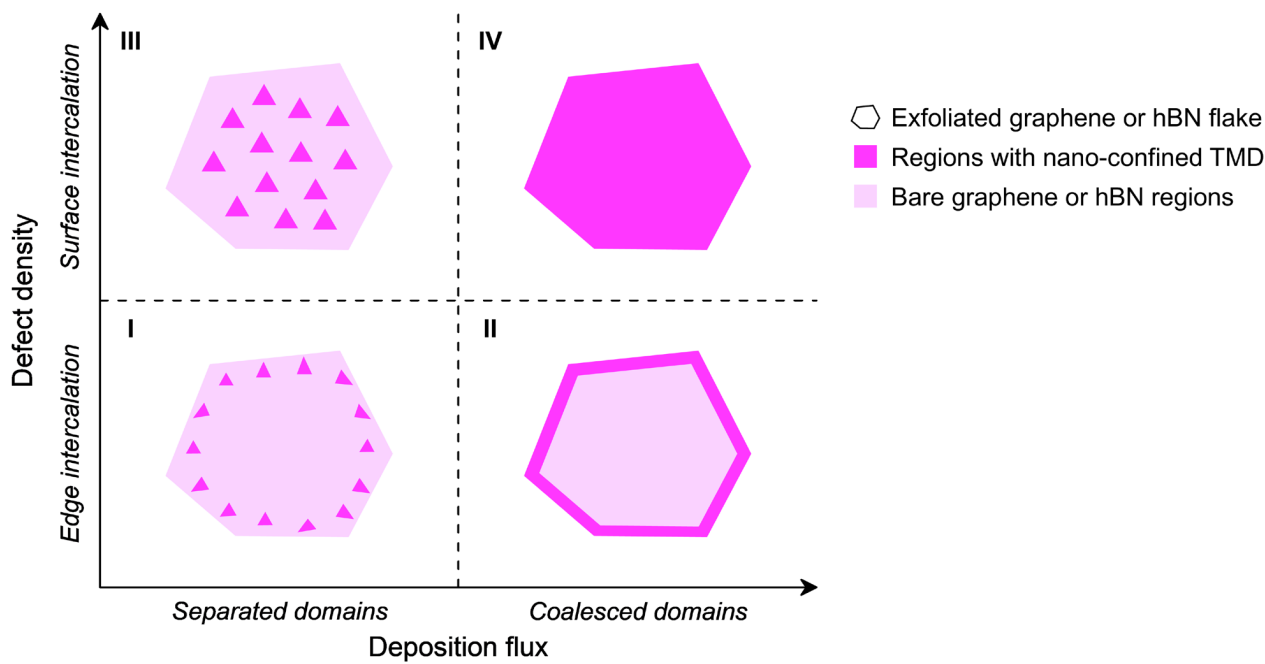

**Fig. S17 | Phase diagram of the nano-confined growth.** By further tunning the deposition flux of precursors and the density of defects in the capping layers, the growth morphology under the nano-confinement exhibits distinct features that can be explained using a four-regime phase diagram: (I) triangular TMD domains with small lateral sizes, preferentially grown near the edges of the capping layers; (II) intrinsically-patterned TMD rings that precisely conform to the edge contours; (III) large triangular TMD domains evenly distributed across the capping layers; (IV) continuous TMD films. When the capping layer is intact and contains a negligible number of defects, the edges serve as the only available paths for intercalation. Under this condition, nucleation preferentially occurs near the edges because of a higher local concentration of precursors. When the deposition flux, controlled by the amount of the transition metal source (Table S1), is maintained at a low level, the nano-confined growth results in small triangular domains along the edges, exhibiting directed overgrowth behavior (Figs. 2b, S2, S3, and S14; Growth Regime I). When a high deposition flux is obtained by increasing the amount of the transition metal source, initially isolated domains coalesce into continuous rings that precisely follow the edge contours, enabling the intrinsically-patterned growth of TMD monolayers (Figs. 4, S19, and S20; Growth Regime II). In addition, the intercalation of precursors is facilitated by defects in the capping layers. For graphene, defects are introduced by high-temperature annealing of exfoliated samples in $O_2$/Ar atmosphere[38]. These defects serve as additional intercalation paths,

allowing surface intercalation of precursors and thereby promoting homogeneous nucleation within the nano-confinement. Under a low deposition flux, the growth process leads to evenly-distributed triangular domains with nearly symmetric shapes and large lateral sizes (Figs. 1b and S18a; Growth Regime III), while a fully-coalesced, continuous TMD film forms under high-flux conditions (Fig. S18b,c; Growth Regime IV). It is evident that the four-regime phase diagram—acting as a navigator for growth morphology—considerably enhances our kinetic control over the nano-confined growth, which enables precise synthesis of tailored TMD monolayers for customized electronic applications.

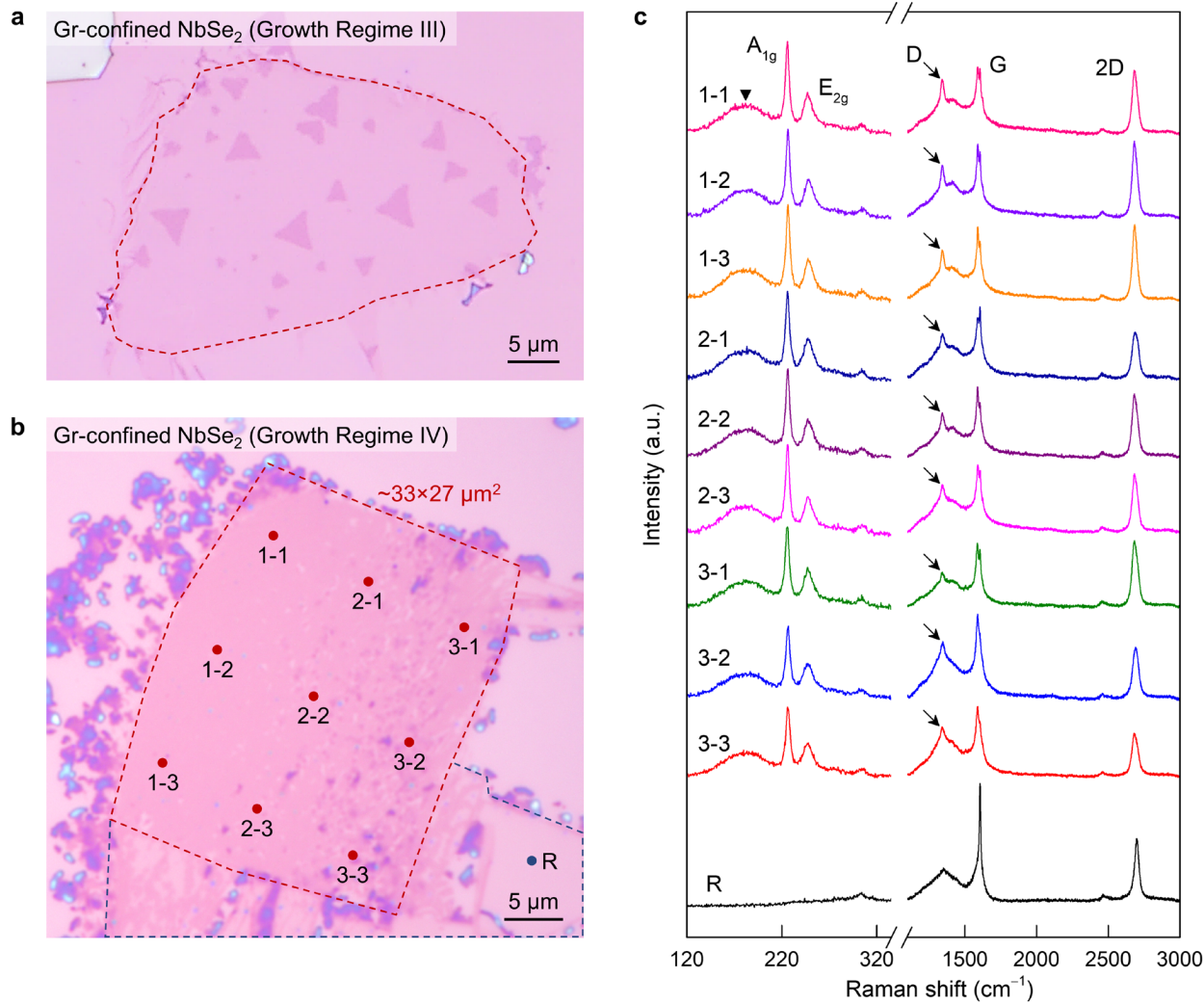

**Fig. S18 | Nano-confined growth of large-scale NbSe₂. a**, Optical micrograph of triangular NbSe$_2$ monolayers underneath a defective graphene flake (marked by red dashed lines), synthesized using a low deposition flux (i.e., Growth Regime III of the phase diagram; Fig. S17). The NbSe$_2$ domains show large lateral sizes and are evenly distributed across the graphene flake, in contrast to the small and edge-localized NbSe$_2$ domains formed underneath intact capping layers (Fig. S2a) (i.e., Growth Regime I). This observation demonstrates that defects serve as additional intercalation paths, which enable surface intercalation of precursors and thereby facilitate homogeneous nucleation within the nano-confinement. **b**, Optical micrograph of a continuous NbSe$_2$ film that fully spans a defective graphene region (indicated by red dashed lines). A high deposition flux was adopted to achieve full coalescence of the NbSe$_2$ monolayer (i.e., Growth Regime IV). The graphene region indicated by blue dashed lines is less defective or even intact, leading to reduced growth of NbSe$_2$ underneath. **c**,

Raman characterization of the sample shown in **b**, comparing the defective (Points 1-1 through 3-3) and intact (Point R) regions. The typical Raman bands of graphene (G and 2D) are clearly visible in both regions; however, the spectral window before the G band exhibits distinct line shapes. A broad feature is observed in the intact region that can be attributed to amorphous carbon contamination[39], whereas the defective region exhibits a narrow D band, an explicit indicator of defects[40], in addition to the broad feature (to aid identification, the D band is marked by arrows). As for $NbSe_2$, its typical Raman bands ($A_{1g}$ and $E_{2g}$) along with the soft mode (indicated by ▼) are consistently observed at all measured positions in the defective region, corroborating the fully-coalesced nature of the $NbSe_2$ monolayer. However, all features are absent at the reference Point R located within the intact region, verifying the absence of $NbSe_2$ growth. Such a strong correlation between defects and morphology offers explicit evidence for the capability of the nano-confined growth to fabricate large-scale TMD monolayers, thereby enabling applications in 2D integrated circuits where high coverage is essential.

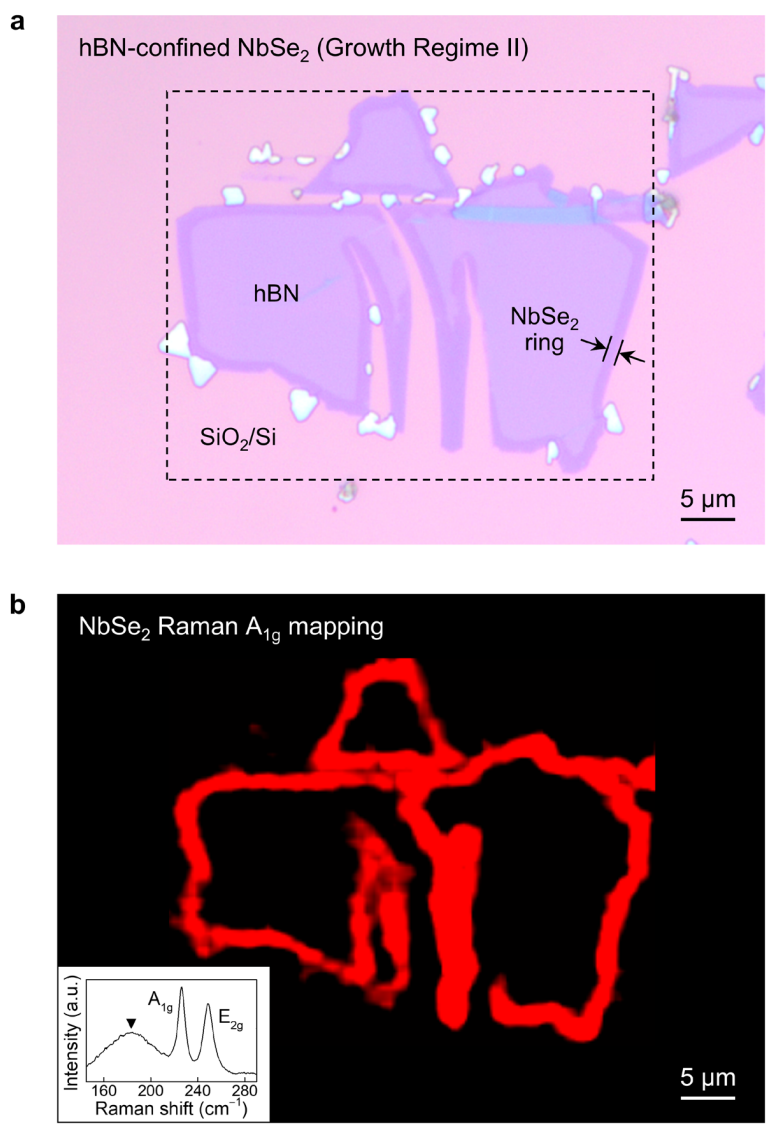

**Fig. S19 | Nano-confined growth of intrinsically-patterned NbSe$_2$. a**, Optical micrograph of intrinsically-patterned NbSe$_2$ monolayer rings with as-exfoliated hBN flakes as the capping layers, synthesized within Growth Regime II of the phase diagram (Fig. S17). It is evident that the NbSe$_2$ rings precisely follow the edge contours of the hBN flakes. **b**, Raman mapping of the dashed region in **a**, which demonstrates the continuity of the NbSe$_2$ rings. Inset: Raman spectrum acquired on the NbSe$_2$ ring, clearly revealing the typical A$_{1g}$ and E$_{2g}$ bands along with the soft mode (marked by ▼).

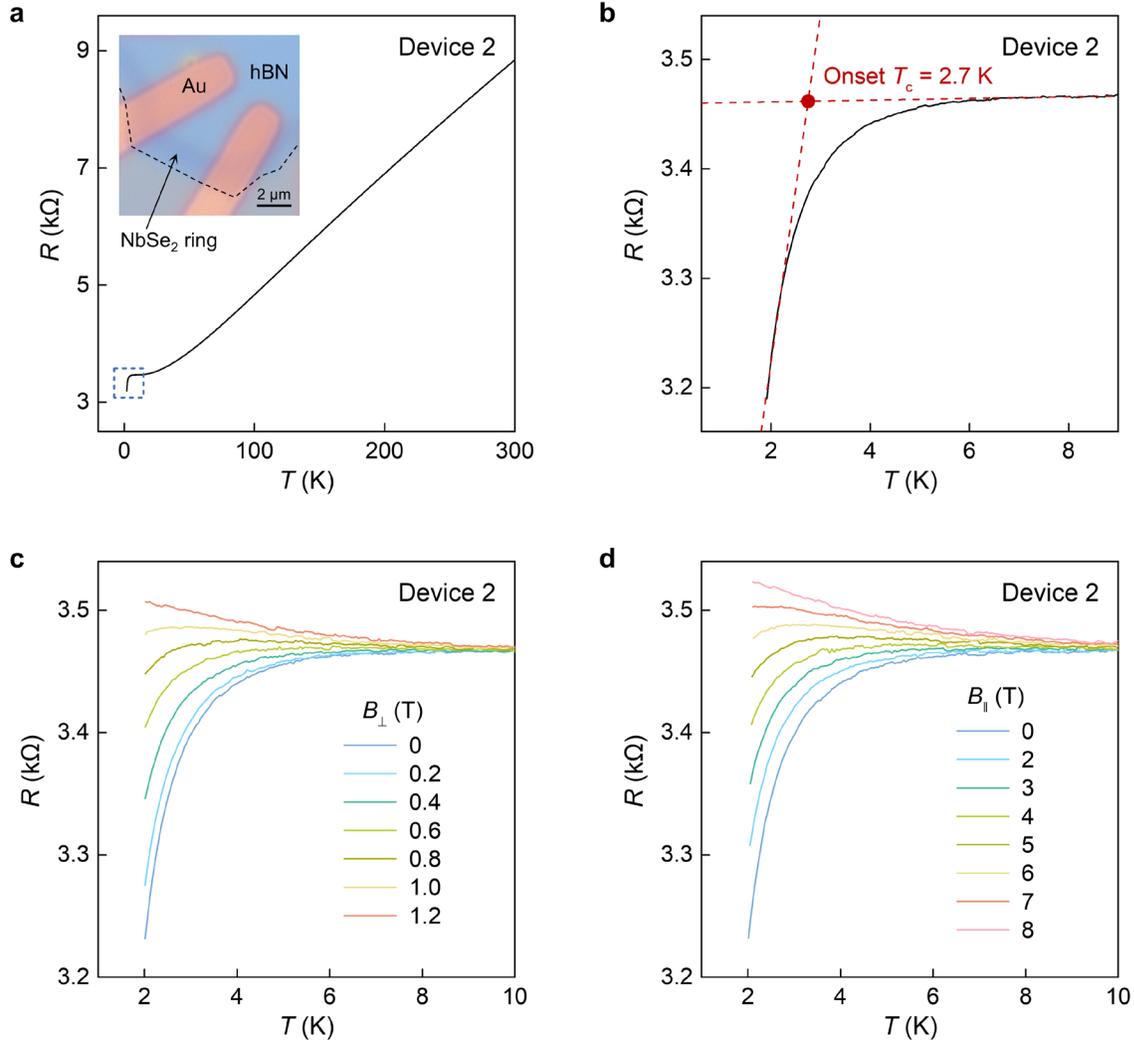

**Fig. S20 | Superconductivity of intrinsically-patterned NbSe$_2$. a**, *R-T* curve of the intrinsically-patterned NbSe$_2$ monolayer, showing linear metallic behavior followed by a sharp superconducting transition. Inset: optical micrograph of the device. Dashed lines indicate hBN edges. **b**, Close-up of the low-*T* region (marked by blue dashed lines in the main panel of **a**), highlighting the onset of the superconducting transition at 2.7 K, in good agreement with the 2.8 K determined for Device 1 (Fig. 4g). **c,d**, *R-T* curves measured under perpendicular (**c**) and parallel (**d**) magnetic fields. As expected, the superconducting transition is quickly suppressed under perpendicular fields but shows enhanced robustness under parallel fields owing to strong Ising pairing protected by spin-orbit interaction[41,42].

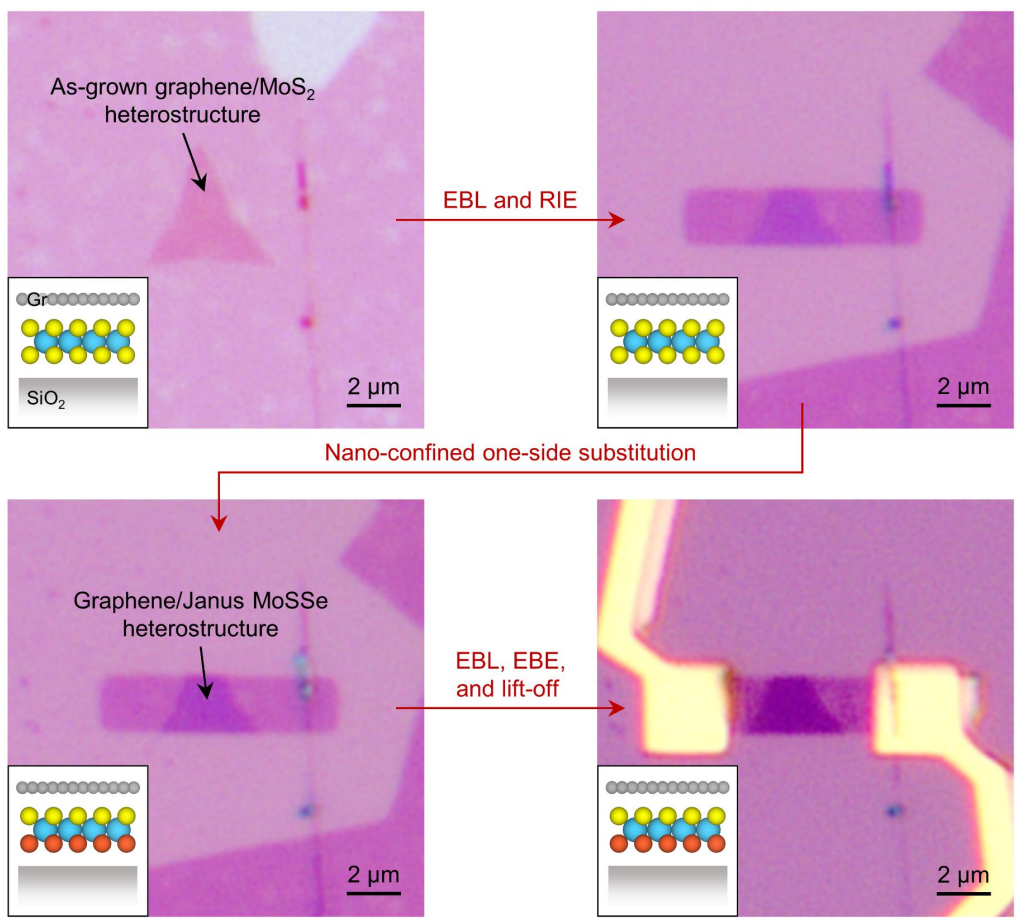

**Fig. S21 | Fabrication of the graphene/Janus MoSSe heterostructure device.** A series of optical micrographs illustrating the fabrication procedure. Insets: cross-sectional schematic diagrams of the sample with the gray, cyan, yellow, and orange balls denoting C, Mo, S, and Se atoms, respectively. First, a bar-shaped channel was patterned on a graphene/MoS$_2$ heterostructure using electron-beam lithography (EBL; Raith eLINE Plus) and O$_2$ reactive-ion etching (RIE; Oxford Instruments Plasma Lab 80Plus), which was followed by one-side substitution to convert MoS$_2$ to Janus MoSSe. Next, electrodes were defined using EBL, followed by electron-beam evaporation (EBE; AST Peva-600E) of a Cr/Au film (5/50 nm) and a lift-off process in acetone. Device measurements were carried out at room temperature in vacuum, using a probe station (Lake Shore) equipped with a 450-nm laser. A precision source/measure unit (Keysight B2902A) was used to apply voltages and measure currents.

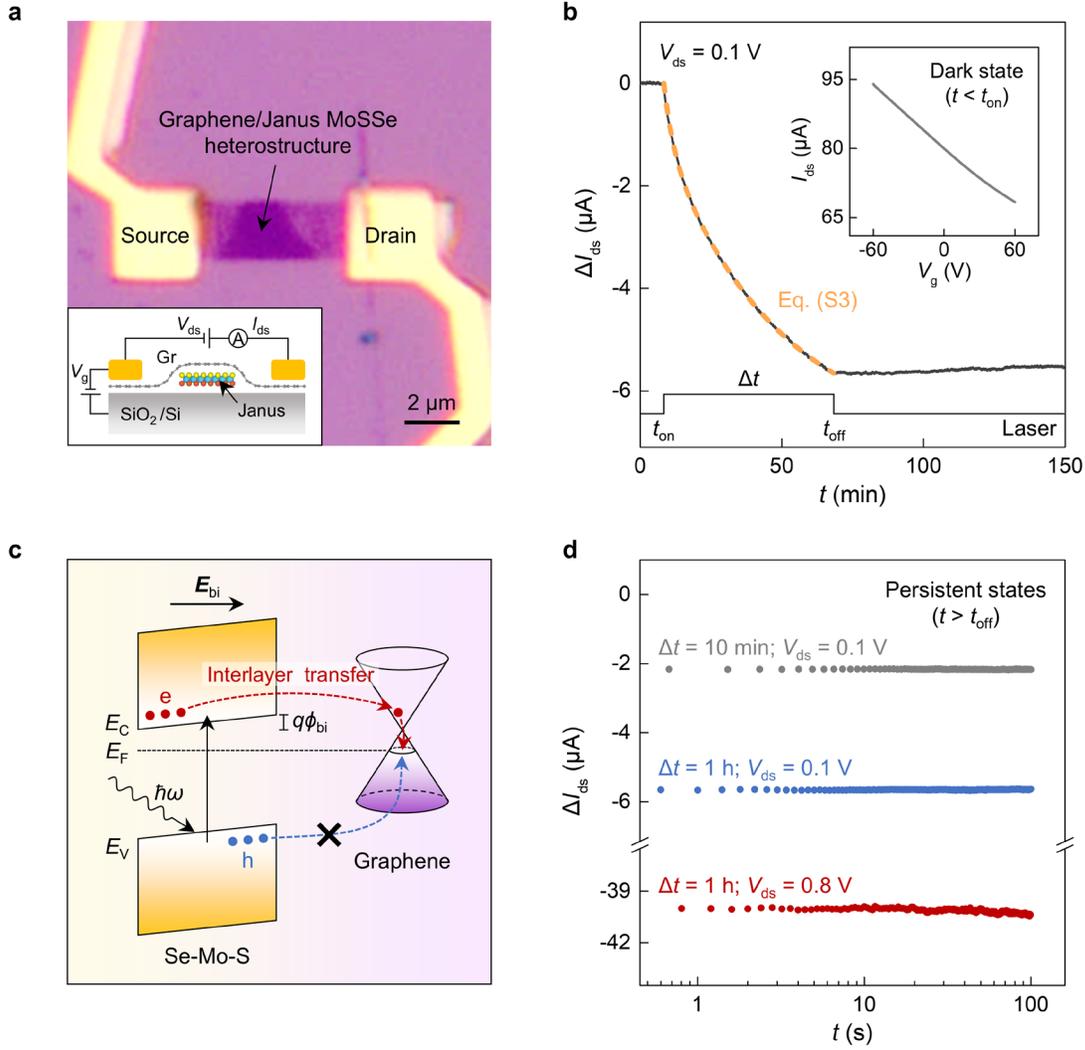

**Fig. S22 | Electronic properties of the graphene/Janus MoSSe heterostructure.**

**a**, Optical micrograph of the device. The graphene/Janus MoSSe heterostructure is located at the center of the bar-shaped channel. The source and drain electrodes are in contact with the graphene flake. Inset: schematic diagram of the device. $I_{ds}$, channel current; $V_{ds}$, bias voltage; $V_g$, gate voltage.

**b**, Photoresponse (defined as $\Delta I_{ds} = I_{ds}^{light} - I_{ds}^{dark}$) under 450-nm laser excitation with $V_{ds}$ = 0.1 V and $V_g$ = 0 V. The bottom line illustrates the on-off switching of the laser. Inset: transfer characteristics of the device under dark conditions with $V_{ds}$ = 0.1 V, exhibiting *p*-type conduction behavior. The weak modulation of current, evidenced by a small max-to-min current ratio of 1.4, implies that the semi-metallic graphene flake serves as the dominant transport channel in the heterostructure[43-45]. As further shown in the main panel, $I_{ds}$ gradually decreases when the laser is turned on at $t_{on}$, indicating an increase in the graphene resistance, which can be attributed to the photodoping effect[44-47]. Under laser excitation, electron-hole pairs are created in the semiconducting Janus MoSSe monolayer. The photoexcited electrons then transfer to the *p*-type graphene channel, which reduces the original hole density and thereby increases the graphene resistance. Moreover, the time dependence of $\Delta I_{ds}$ can be well described using the double-exponential function (orange dashed curve)

$$\Delta I_{ds} = I_1\left(1 - e^{-\frac{t-t_{on}}{\tau_1}}\right) + I_2\left(1 - e^{-\frac{t-t_{on}}{\tau_2}}\right), \tag{S3}$$

where $I_1$ and $I_2$ are the saturated contributions, while $\tau_1$ and $\tau_2$ are the time constants. The fitting yields a faster $\tau_1 = 203$ s and a slower $\tau_2 = 2443$ s, which are attributed to the intrinsic and defect-mediated electron-transfer processes[44,48], respectively. We note that the intrinsic $\tau_1$ is two orders of magnitude larger than those reported for graphene/MoS$_2$ devices (e.g., $\tau_1 \sim 2$ s in ref. [44] and $\tau_1 < 10$ s in ref. [45]), which indicates a reduced electron transfer rate in our device, likely arising from the polar chalcogen arrangement of the Janus MoSSe monolayer. Meanwhile, the extrinsic $\tau_2$ may correlate with defects and indicate the presence of charge traps. Given the superior quality of the Janus MoSSe crystal as revealed by the narrow Raman and PL linewidths (Fig. S23) and its ultraclean vdW interface with graphene as confirmed by cross-sectional STEM (Fig. S15), it is likely that the charge traps mainly originate from the Janus MoSSe/SiO$_2$ interface. This is consistent with the unsaturated SiO$_2$ surface which contains a large number of dangling bonds that interact with the Janus MoSSe monolayer[49,50].

**c**, Energy-band diagram of the graphene/Janus MoSSe heterostructure showing the charge-transfer mechanism at the vdW interface. The Janus MoSSe monolayer is *n*-type based on the literature[10,11]. $E_F$, Fermi level; $E_C$, conduction band edge; $E_V$, valence band edge. As illustrated in the energy-band diagram, the polar arrangement of the Janus MoSSe monolayer leads to a built-in potential ($\phi_{bi}$) and an associated electric field ($E_{bi}$) within the monolayer[51,52]. Under laser excitation, the photoexcited electrons and holes become spatially separated, with the electrons moving toward the Se side with a lower potential energy[53,54]. The coupling of wavefunctions between this "biased" donor state and the acceptor state in graphene is considerably reduced due to an extended donor-acceptor distance, which results in the reduced transfer rate of the photoexcited electrons and the long time constant[55]. Another factor that controls the charge-transfer process is the energy difference between the donor and acceptor states, as an efficient transfer occurs only when the two states are nearly degenerate[55]. As illustrated in the energy-band diagram, the graphene Dirac cone aligns with the $E_C$ of the Janus MoSSe monolayer, yielding a series of electron donor and acceptor states with degenerate energies. The situation is different for the hole donor states near the $E_V$ because of the absence of degenerate acceptor states in the Dirac cone[53]. Therefore, the hole transfer across the vdW interface (blue arrow) is hard to realize, while the electron transfer (red arrow) governs the photoresponse.

**d**, Manipulation of persistent states. As shown in the main panel of **b**, when the laser is turned off at $t_{off}$, the device exhibits a persistent photoresponse with $\Delta I_{ds}$ retaining ~98% of its initial magnitude even after ~1.4 h. Here, by further tunning the illumination time ($\Delta t$) and the bias voltage ($V_{ds}$) with $V_g = 0$ V, a series of persistent states can be achieved with different levels of $\Delta I_{ds}$. Such a persistent behavior reflects the ultralong lifetime of the charge-transfer state, which may result from the slow back electron-transfer process[56] and the slow exciton recombination in the Janus MoSSe monolayer due to the spatial separation of the electron and hole wavefunctions[57]. The above findings highlight the potential applications of the graphene/Janus MoSSe heterostructure in 2D photomemory devices.

# Part III | Comparisons with literature

**Table S8 | Comparison with reported nano-confined growth techniques.** Under the "Defective capping layer" column, # indicates defects introduced by plasma treatment; * indicates intercalation through pre-existing point defects or tears in the capping layers; † indicates defects arising from interactions between the capping layers and the precursors. In contrast, the edge-intercalation mechanism proposed in our study provides a distinct route to achieve nano-confined growth without relying on defective capping layers, enabling intrinsic-patterning capability through further tuning of growth kinetics (Fig. S17). Moreover, our method represents the only nano-confined technique capable of synthesizing Janus 2D materials with polar structures, and the only technique that enables fully insulating encapsulation with hBN as the capping layer rather than conductive graphene. These advancements clearly distinguish our approach from existing techniques in the field.

| Reference | Nano-confinement | Intercalated materials | Defective capping layer | Intrinsic patterning | Synthesis of Janus materials | Fully insulating encapsulation |
|---|---|---|---|---|---|---|
| This work | Gr/SiO$_2$ or hBN/SiO$_2$ | TMDs | Not necessary | Yes | Yes | Yes (with hBN) |
| Briggs et al.[58] | Gr/SiC | Ga, In, and Sn | Required[#] | No | No | No |
| Al Balushi et al.[59] | Gr/SiC | GaN | Required[*] | No | No | No |
| Pécz et al.[60] | Gr/SiC | InN | Required[†] | No | No | No |
| Wang et al.[61] | Gr/Si | AlN | Required[#] | No | No | No |
| Li et al.[62,63] | Gr/Ru | Silicene | Required[†] | No | No | No |

**Table S9 | Comparison with reported CVD techniques for NbSe$_2$ growth.** All reported techniques utilize open growth environments. It is clear that our method demonstrates substantial advancements in the synthesis of NbSe$_2$, achieving enhanced monolayer uniformity, superior crystal quality, and exceptional air stability, all of which contribute to its excellent superconducting performance. Moreover, our method enables various growth morphologies—suitable for applications ranging from large-scale integration to superconducting circuits.

| Reference | Monolayer yield | Air stability | Monolayer onset $T_c$ | Growth morphology |
|---|---|---|---|---|
| This work | 98% | Monolayers stable in air for over 60 days | 2.8 K | Domains, films, or patterned rings |
| Wang et al.[3] | 81% | Monolayers only stable in vacuum or with post-synthesis encapsulation | 1.2 K | Domains |
| Park et al.[64] | 77% | Monolayers partially oxidized in air | No data | Domains |
| Zhou et al.[14] | Monolayer applicable (yield not specified) | No data | 1.4 K | Domains |
| Lin et al.[65] | Technique limited to bilayers | Bilayers stable in air for over 10 days | No data | Films |
| Zou et al.[66] | Technique limited to thick nanoplates | No data | No data | Domains |

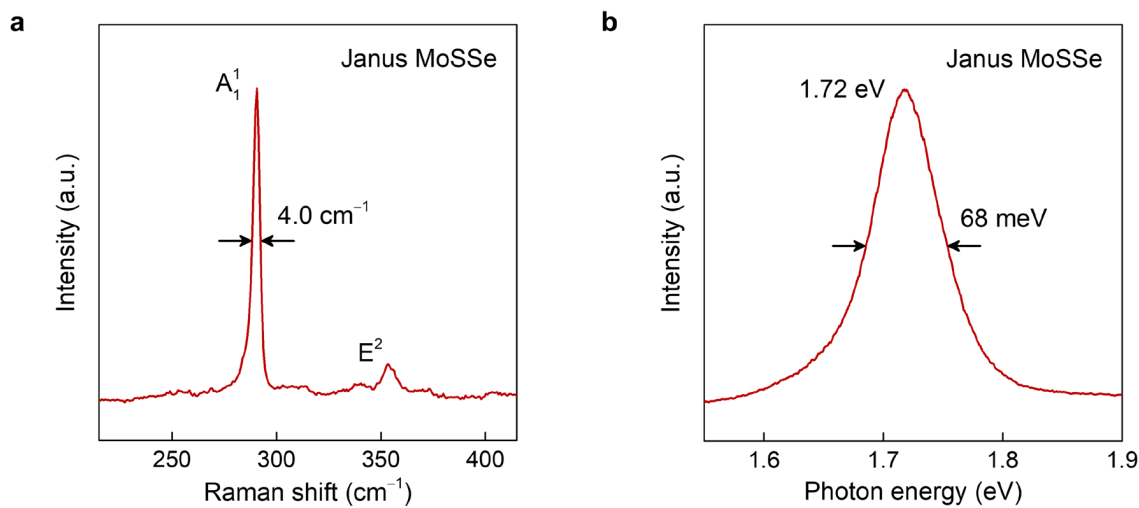

| Raman $A_1^1$ band | | PL peak | | Reference |
|---|---|---|---|---|
| Position (cm⁻¹) | FWHM (cm⁻¹) | Position (eV) | FWHM (meV) | |
| 290 | 4.0 | 1.72 | 68 | This work |
| 290 | 5.3 | 1.70 | 80 | Liu et al. |
| 287 | 7.3 | 1.66 | 75 | Suzuki et al. |
| 290 | 4.5 | 1.72 | 61* | Gan et al. |
| 288 | 5.7 | 1.72 | 121 | Guo et al. |
| 290 | 5.3 | 1.79 | 90 | Zheng et al. |
| 290 | 10.3 | 1.68 | 91 | Trivedi et al. |
| 290 | 7.2 | 1.68 | 88 | Zhang et al. |
| 288 | 8.6 | 1.68 | 91 | Lu et al. |

* The sample was sandwiched by hBN after synthesis to reduce the FWHM of the PL peak.

**Fig. S23 | Comparison of quality benchmarks for Janus MoSSe monolayers. a,b**, Raman (**a**) and PL (**b**) spectra of hBN-confined Janus MoSSe monolayers collected using 1800 and 600 gr/mm gratings, respectively. The FWHMs of Raman and PL peaks are commonly used as benchmarks for assessing the crystal quality of 2D materials[67]. To compare the quality of our Janus MoSSe monolayers with those synthesized via alternative substitution techniques, the FWHMs were extracted from **a** and **b** (marked by arrows). **c**, Comparison of the Raman and PL FWHMs with the literature[9-13,57,68,69]. Our Janus MoSSe monolayers exhibit notably narrower Raman and PL linewidths, demonstrating superior crystal quality. The key advantage of our method is the incorporation of a vdW capping layer (i.e., hBN or graphene), which effectively protects the top chalcogen atoms from unintentional modification (Notes S2 and S3), thereby enabling the atomically-precise synthesis of high-quality Janus TMD monolayers.